\documentclass[12pt]{article}
\usepackage{verbatim,epsfig,amsmath,chicago,url}
\newcommand{\E}{\mbox{E}}

\newcommand{\SD}{\mbox{SD}}

\newcommand{\MSE}{\mbox{MSE}}

\newcommand{\ISE}{\mbox{ISE}}

\def\ra{\rightarrow}

\def\beqn{\begin{eqnarray*}}
\def\eeqn{\end{eqnarray*}}
\def\beq{\begin{eqnarray}}
\def\eeq{\end{eqnarray}}
\def\cv{cross-validation~}

\begin{document}
\title{\bf An Empirical Study of Indirect Cross-validation}
\author{Olga Y. Savchuk, Jeffrey D. Hart, Simon J. Sheather}
%\date{\today}
\date{}
\maketitle

\begin{abstract}

In this paper we provide insight into the empirical properties of
indirect cross-validation (ICV), a new method of bandwidth selection
for kernel density estimators. First, we describe the method and
report on the theoretical results used to develop a
practical-purpose model for certain ICV parameters. Next, we provide
a detailed description of a numerical study which shows that the ICV
method usually outperforms least squares cross-validation (LSCV) in
finite samples. One of the major advantages of ICV is its increased
stability compared to LSCV. Two real data examples show the benefit
of using both ICV and a local version of ICV.

\vspace{1cm}

\noindent KEY WORDS: Cross-validation; Bandwidth selection; Kernel
density estimation, Integrated Squared Error, Mean Integrated
Squared Error.
\end{abstract}

\section{Introduction}

Let $X_1,\ldots,X_n$ be a random sample from an unknown density $f$.
A kernel density estimator of $f$ at the point $x$ is defined as
\begin{equation}
\label{eq:KDE} \hat f_h(x)=\frac{1}{nh}\sum_{i=1}^n
K\Bigl(\frac{x-X_i}{h}\Bigr),
\end{equation}
where $h>0$ is the bandwidth, and $K$ is the kernel, which is
generally chosen to be a unimodal probability density function that
is symmetric about zero and has finite variance.  A popular choice
for $K$ is the Gaussian kernel: $\phi(u)=(2\pi)^{-1/2}\exp(-u^2/2)$.
To distinguish between estimators with different kernels, we shall
refer to estimator \eqref{eq:KDE} with given kernel $K$ as a {\it
$K$-kernel estimator}.

Practical implementation of the estimator~\eqref{eq:KDE} requires
specification of the smoothing parameter $h$. The two most widely
used bandwidth selection methods are least squares cross-validation,
proposed independently by~\citeN{Rudemo:LSCV}
and~\citeN{Bowman:LSCV}, and the~\citeN{Sheather:PI} plug-in method.
Plug-in is often preferred since it produces more stable bandwidths
than does LSCV. Nevertheless, the LSCV method is still popular since
it requires fewer assumptions than the plug-in method and works well
when the density is difficult to estimate;
see~\citeN{Loader:Classical}, ~\citeN{vanEs},
and~\citeN{Sainetal:cv}.

The main flaw of LSCV is high variability of the selected
bandwidths. Other drawbacks include the tendency of cross-validation
curves to exhibit multiple local minima with the first local minimum
being too small (see ~\citeN{HallMarron:LocMin}), and the tendency
of LSCV to select bandwidths that are much too small when the data
exhibit a small amount of autocorrelation
(see~\citeN{Hart:Autocorrelation} and~\citeN{Cao:Dependence} for
results of a numerical study). Many modifications of LSCV have been
proposed in an attempt to improve its performance. These include
biased cross-validation of ~\citeN{Scott:UCV}, a method
of~\citeN{Chiu:CV}, the trimmed cross-validation
of~\citeN{Feluch:spacing}, the modified cross-validation
of~\citeN{Stute}, and the method of~\citeN{Ahmad} based on kernel
contrasts.

This paper is concerned with a new modification of the LSCV method,
called {\it indirect cross-validation} (ICV), recently proposed by
the authors~\citeN{SavchukHartSheather:ICV}. The ICV method depends
on two parameters, $\alpha$ and $\sigma$. A main theoretical result
is that at asymptotically optimal choices of $\alpha$ and $\sigma$
the ICV bandwidth can converge to zero at a rate $n^{-1/4}$, which
is substantially better than the $n^{-1/10}$ rate of LSCV. The
present paper contains the results of an empirical study of ICV. In
Section~\ref{sec:description} we provide a description of the
method. Section~\ref{sec:Practical} contains the details underlying
the development of a practical purpose model for $\alpha$ and
$\sigma$. Section~\ref{sec:Sims} outlines the results of a numerical
study which, in particular, show that ICV has greater stability in
finite samples than does LSCV. In Section~\ref{sec:Data} we apply
ICV and a local version of ICV to real data sets.
Section~\ref{sec:Summary} provides a summary of our results.

\section{Description of indirect cross-validation\label{sec:description}}

\subsection{Notation and definitions}
We begin with some notation and definitions that will be used
subsequently. For an arbitrary function $g$, define
\[
R(g)=\int g(u)^2\,du,\quad \mu_{jg}=\int u^j g(u)\,du,
\]
where here and subsequently integrals are assumed to be over the
whole real line. The popular measures of performance of the kernel
estimators~\eqref{eq:KDE} are integrated squared error (ISE) and
mean integrated squared error (MISE). The ISE is defined as
\begin{equation}
\label{eq:ISE} ISE(h)=\int \bigl(\hat{f}_h(x)-f(x)\bigr)^2\,dx,
\end{equation}
and MISE is defined as the expectation of ISE. Assuming that the
underlying density $f$ has second derivative which is continuous and
square integrable and that $R(K)<\infty$, the bandwidth which
asymptotically minimizes the MISE of the $K-$kernel
estimator~\eqref{eq:KDE} has the following form:
\begin{equation}
\label{eq:h_n}
h_n=\left\{\frac{R(K)}{\mu_{2K}^2 R(f^{\prime
\prime})}\right\}^{1/5}n^{-1/5}.
\end{equation}

The LSCV criterion is given by
\begin{equation}
\label{eq:LSCV} LSCV(h)=R(\hat f_h)-\frac{2}{n}\sum_{i=1}^n \hat
f_{h,-i}(X_i),
\end{equation}
where $\hat f_{h,-i}$ denotes the kernel estimator (\ref{eq:KDE})
constructed from the data without the observation $X_i$. A well
known fact is that $LSCV(h)$ is an unbiased estimator of
$MISE(h)-\int f^2(x)\,dx$. For this reason the LSCV method is often
called {\it unbiased cross-validation}. Let $\hat h_{UCV}$ and $h_0$
denote the bandwidths which minimize the LSCV
function~\eqref{eq:LSCV} and the MISE of the $\phi$-kernel
estimator. Section~\ref{sec:Method} defines the ICV bandwidth,
denoted as $\hat h_{ICV}$.

\subsection{The basic method\label{sec:Method}}

The essence of the ICV method is to use different kernels at the
cross-validation and density estimation stages. The same idea is
exploited by the one-sided cross-validation method of~\citeN{HartYi}
in the regression context. ICV first selects the bandwidth of an
$L-$kernel estimator using least squares cross-validation. Selection
kernels $L$ used for this purpose are described in
Section~\ref{sec:Kernels}. The bandwidth so obtained is rescaled so
that it can be used with the $\phi$-kernel estimator. The
multiplicative constant $C$ has the following form:
\begin{equation}
\label{eq:C}
C=\left(\frac{\mu_{2L}^2}{2\sqrt{\pi}R(L)^2}\right)^{1/5},
\end{equation}
which is motivated by the asymptotically optimal MISE
bandwidth~\eqref{eq:h_n}.

\subsection{Selection kernels \label{sec:Kernels}}

We consider the family of kernels ${\cal
  L}=\{L(\,\cdot\,;\alpha,\sigma): \alpha\ge0,\sigma>0\}$, where, for
all $u$,
\begin{equation}
\label{eq:L}
L(u;\alpha,\sigma)=(1+\alpha)\phi(u)-\frac{\alpha}{\sigma}\phi\left(\frac{u}{\sigma}\right).
\end{equation}
Note that the Gaussian kernel is a special case of~\eqref{eq:L} when
$\alpha=0$ or $\sigma=1$. Each member of ${\cal L}$ is symmetric
about 0 and has the second moment $\mu_{2L}=\int
u^2L(u)\,du=1+\alpha-\alpha\sigma^2$. It follows that kernels in
${\cal L}$ are second order, with the exception of those for which
$\sigma=\sqrt{(1+\alpha)/\alpha}$.

\begin{figure}
\begin{center}
\begin{tabular}{cc}
{\bf(a)}&{\bf(b)}\\
\epsfig{file=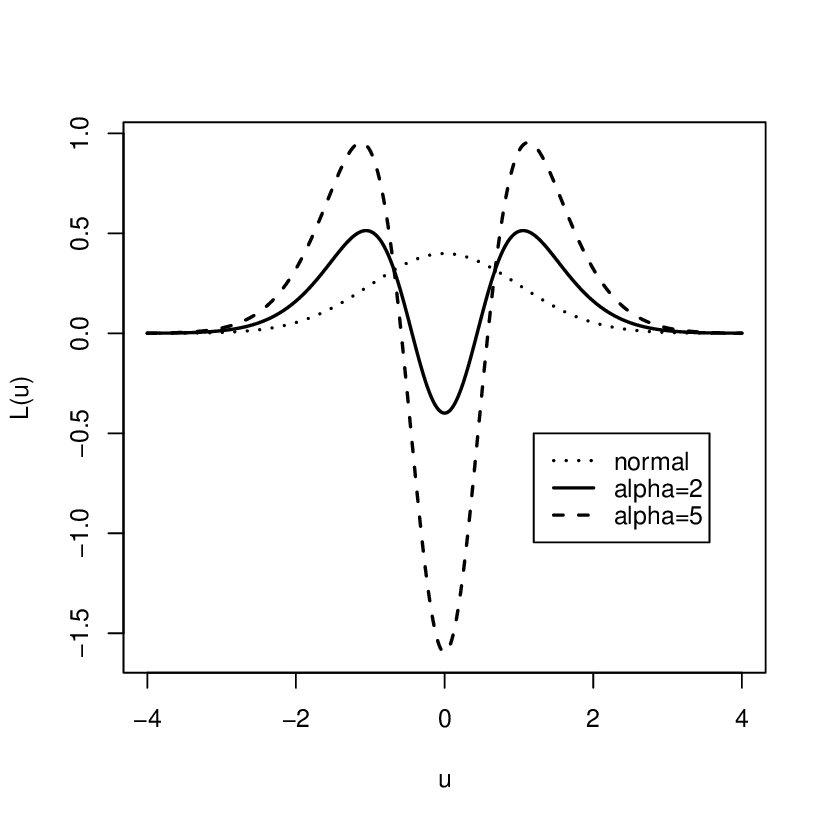,height=220pt}&\epsfig{file=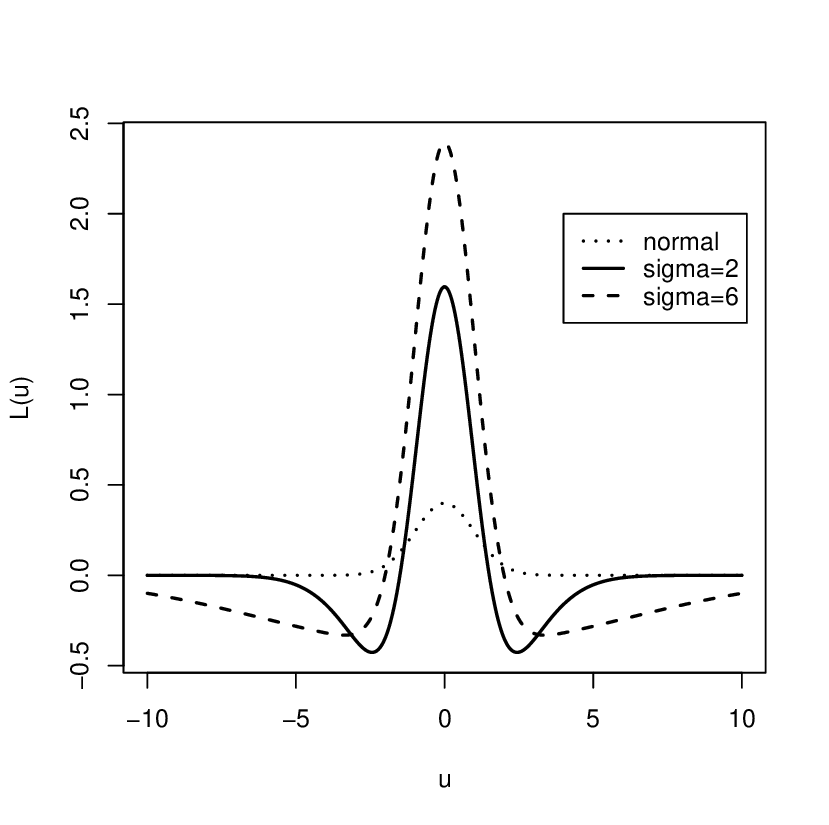,height=220pt}\\
\end{tabular}
\caption{{\bf(a)} Selection kernels in $\mathcal L_1$ which have
$\sigma=0.5$; {\bf(b)} Selection kernels in $\mathcal L_3$ with
$\alpha=6$. The dotted curve in both graphs corresponds to the
Gaussian kernel.\label{fig:Lplot}}
\end{center}
\end{figure}

The family ${\cal L}$ can be partitioned into three families: ${\cal
L}_1$,  ${\cal L}_2$ and  ${\cal L}_3$. The first of these is
$\mathcal L_1=\bigl\{L(\cdot;\alpha,\sigma): \alpha>0,
\sigma<\frac{\alpha}{1+\alpha}\bigr\}$. Each kernel in ${\cal
  L}_1$ has a negative dip centered at $x=0$. The kernels in $\mathcal L_1$ are ones that
``cut-out-the-middle,'' some examples of which are shown in
Figure~\ref{fig:Lplot}{\bf (a)}.

The second family is $\mathcal L_2=\bigl\{L(\cdot;\alpha,\sigma):
\alpha>0, \frac{\alpha}{1+\alpha}\leq\sigma\leq1\bigr\}$. Kernels in
$\mathcal L_2$ are densities which can be unimodal or bimodal. Note
that the Gaussian kernel is a member of this family. The third
family is $\mathcal L_3=\bigl\{L(\cdot;\alpha,\sigma): \alpha>0,
\sigma>1\}$, each member of which has negative tails. Examples are
shown in Figure~\ref{fig:Lplot}{\bf (b)}.

Kernels in $\mathcal L_1$ and $\mathcal L_3$  turn out to be highly
efficient for \cv purposes but very inefficient for estimating $f$.
This explains why we do not use $L$ as both a selection {\it and} an
estimation kernel.

Selection kernels in $\cal L$ are mixtures of two normal densities,
which greatly simplifies computations. In particular, closed form
expressions exist for the $LSCV$ and $\ISE$ functions. This fact has
been utilized by ~\citeN{MarronWand:MISE} to derive exact MISE
expressions. \citeN{MarronWand:MISE} point out that, in addition to
their computational advantages, normal mixtures can approximate any
density arbitrarily well in various senses. Mixtures of normals are
therefore an excellent model for use in simulation studies, a fact
which we take advantage of in Section \ref{sec:Sims}.

\section{Practical issues\label{sec:Practical}}

In this section we address the problem of choosing the parameters,
$\alpha$ and $\sigma$, of the selection kernel in practice. We
review some large sample theory for the ICV method and provide the
theoretical results used to develop the practical-purpose model for
$\alpha$ and $\sigma$.

\subsection{Large sample theory \label{sec:LargeSample}}

Large sample theory was developed in~\citeN{SavchukHartSheather:ICV}
by considering the asymptotic mean squared error (MSE) of the ICV
bandwidth. Their results may be summarized as follows.

\begin{enumerate}
\item Under suitable regularity conditions the ICV bandwidth is
asymptotically normally distributed.

\item The asymptotic MSE of $\hat h_{ICV}$ has been found for two cases: $\sigma\ra
0$ (cut-out-the-middle kernels) and $\sigma\ra\infty$
(negative-tailed kernels). It turns out that when the asymptotically
optimal values of $\alpha$ and $\sigma$ are used in the respective
cases, the MSE converges to zero at the same rate of $n^{-9/10}$,
but the limiting ratio of optimum mean squared errors is $0.752$,
with $\sigma\ra\infty$ yielding the smaller error. In comparison,
the rate at which the MSE for $\hat h_{UCV}$ converges to zero is
$n^{-6/10}$.

The subsequent theoretical results are provided for the case
$\sigma\ra\infty$.

\item The relative rate of convergence of $\hat h_{ICV}$ to $h_0$ is
$n^{-1/4}$, whereas the corresponding rate for $\hat h_{UCV}$ is
$n^{-1/10}$.

\item Values of $\sigma$ which minimize the asymptotic MSE are as
follows:
\begin{equation}
\label{eq:sigopt}
\sigma_{n,opt}=n^{3/8}A_\alpha\left[\frac{R(f)R(f'')^{13/5}}{R(f''')^2}\right]^{5/8},
\end{equation}
where $A_\alpha=\displaystyle{16\sqrt\pi\frac{
2^{7/16}}{3^{5/8}}\frac{\alpha^{3/4}}{(1+\alpha)^2}\left(\frac{1}{8}(1+\alpha)^2-\frac{8}{9\sqrt3}(1+\alpha)+\frac{1}{\sqrt2}\right)^{5/8}}$.
\item The asymptotically optimal $\alpha$ is 2.4233.
Remarkably, the optimal $\alpha$ does not depend on $f$.

\item When the asymptotically optimal values of $\alpha$ and
$\sigma$ are used, the asymptotic bias and standard deviation of
$\hat h_{ICV}$ converge to zero at the same rate of $n^{-9/20}$.
\end{enumerate}

\subsection{MSE-optimal $\alpha$ and $\sigma$\label{sec:MSEopt}}

Asymptotic results are not always reliable for practical purposes.
In order to have an idea of whether the negative-tailed or
cut-out-the middle kernels should really be used, and how good
choices of $\alpha$ and $\sigma$ vary with $n$ and $f$, we
considered the following expression for the asymptotic MSE of the
ICV bandwidth:
\begin{equation}
\label{eq:MSE}
\begin{array}{ll}
\MSE(\hat h_{ICV})=&\displaystyle{\left(\frac{1}{4\pi}\right)^{1/5}\frac{R(f^{\prime\prime\prime})^2}{R(f^{\prime\prime})^{16/5}}n^{-3/5}\left\{\frac{2}{25}\frac{R(f)R(f^{\prime\prime})^{13/5}}{R(f^{\prime\prime\prime})^2}\frac{R(\rho_L)}{R(L)^{9/5}(\mu_{2L}^2)^{1/5}}+\right.}\\
&\\
&\displaystyle{\left.\frac{n^{-3/5}}{400}\left(\frac{R(L)^{2/5}\mu_{2L}\mu_{4L}}{(\mu_{2L}^2)^{7/5}}-\frac{3}{(4\pi)^{1/5}}
\right)^2\right\}}.
\end{array}
\end{equation}
Expression~\eqref{eq:MSE} is valid for either large or small values
of $\sigma$ and includes second order bias terms.

As our target densities we considered the following five normal
mixtures defined in the article by~\citeN{MarronWand:MISE}:

\begin{center}
\begin{tabular}{ll}
Gaussian density:&$N(0,1)$\\
Skewed unimodal density:&$\frac{1}{5}N(0,1)+\frac{1}{5}N\Bigl(\frac{1}{2},\bigl(\frac{2}{3}\bigr)^2\Bigr)+\frac{3}{5}N\Bigl(\frac{13}{12},\bigl(\frac{5}{9}\bigr)^2\Bigr)$\\
Bimodal
density:&$\frac{1}{2}N\Bigl(-1,\bigl(\frac{2}{3}\bigr)^2\Bigr)+\frac{1}{2}N\Bigl(1,\bigl(\frac{2}{3}\bigr)^2\Bigr)$\\
Separated bimodal
density:&$\frac{1}{2}N\Bigl(-\frac{3}{2},\bigl(\frac{1}{2}\bigr)^2\Bigr)+\frac{1}{2}N\Bigl(\frac{3}{2},\bigl(\frac{1}{2}\bigr)^2\Bigr)$\\
Skewed bimodal
density:&$\frac{3}{4}N(0,1)+\frac{1}{4}N\Bigl(\frac{3}{2},\bigl(\frac{1}{3}\bigr)^2\Bigr)$.\\
\end{tabular}
\end{center}
\begin{table}
\begin{center}
\begin{tabular}{|c|c|c||c|c||c|c||c|c||c|c|}
\hline &\multicolumn{10}{|c|}{Density}\\
\cline{2-11} &\multicolumn{2}{|c||}{}&\multicolumn{2}{|c||}{skewed}&\multicolumn{2}{|c||}{}&\multicolumn{2}{|c||}{separated}&\multicolumn{2}{|c|}{skewed}\\
&\multicolumn{2}{|c||}{normal}&\multicolumn{2}{|c||}{unimodal}&\multicolumn{2}{|c||}{bimodal}&\multicolumn{2}{|c||}{bimodal}&\multicolumn{2}{|c|}{bimodal}\\
\cline{2-11}$n$&$\alpha$&$\sigma$&$\alpha$&$\sigma$&$\alpha$&$\sigma$&$\alpha$&$\sigma$&$\alpha$&$\sigma$\\
\hline 100&3.05&2.79&5.28&1.68&109.68&1.03&16.70&1.19&343.74&1.01\\
\hline 250&2.78&4.04&3.16&2.60&48.46&1.06&4.51&1.84&177.15&1.02\\
\hline 500&2.73&4.97&2.84&3.56&6.21&1.55&3.18&2.58&161.39&1.02\\
\hline 1000&2.69&5.97&2.75&4.49&3.73&2.12&2.84&3.54&123.78&1.03\\
\hline 5000&2.61&8.84&2.66&6.85&2.77&4.26&2.70&5.74&4.71&1.79\\
\hline 20000&2.55&12.40&2.59&9.58&2.68&6.22&2.63&8.08&2.85&3.46\\
\hline 100000&2.50&18.80&2.53&14.27&2.60&9.19&2.56&11.94&2.70&5.65\\
\hline 500000&2.47&29.54&2.49&21.88&2.54&13.65&2.50&18.07&2.62&8.39\\
\hline
\end{tabular}
\caption{MSE-optimal $\alpha$ and $\sigma$.\label{tab:MSE_min}}
\end{center}
\end{table}
These choices for $f$ represent density shapes that are common in
practice.

%It is worth noting that the asymptotically optimal $\sigma$
%(expression~\eqref{eq:sigopt}) is free of location and scale. It is
%thus sensible to choose a single representative of a location-scale
%family when investigating the effect of $f$.

In Table~\ref{tab:MSE_min} we provide the MSE-optimal choices of
$\alpha$ and $\sigma$ for the target densities at eight sample sizes
ranging from $n=100$ up to $n=500000$. It is obvious that the
MSE-optimal $\alpha$ and $\sigma$ vary greatly from one density to
another, which is especially true for ``small'' sample sizes.
However, the optimal $\alpha$ seems to converge to about 2.5 for
each density as $n$ increases, which fits with our observation that
the optimal $\alpha$ is $2.4233$.  The optimal $\sigma$ is
increasing with sample size. It us remarkable that all the
MSE-optimal $\alpha$ and $\sigma$ in Table~\ref{tab:MSE_min}
correspond to kernels from $\mathcal L_3$, the family of
negative-tailed kernels.

\subsection{Model for the ICV parameters}

\begin{figure}
\begin{center}
\begin{tabular}{c}
\epsfig{file=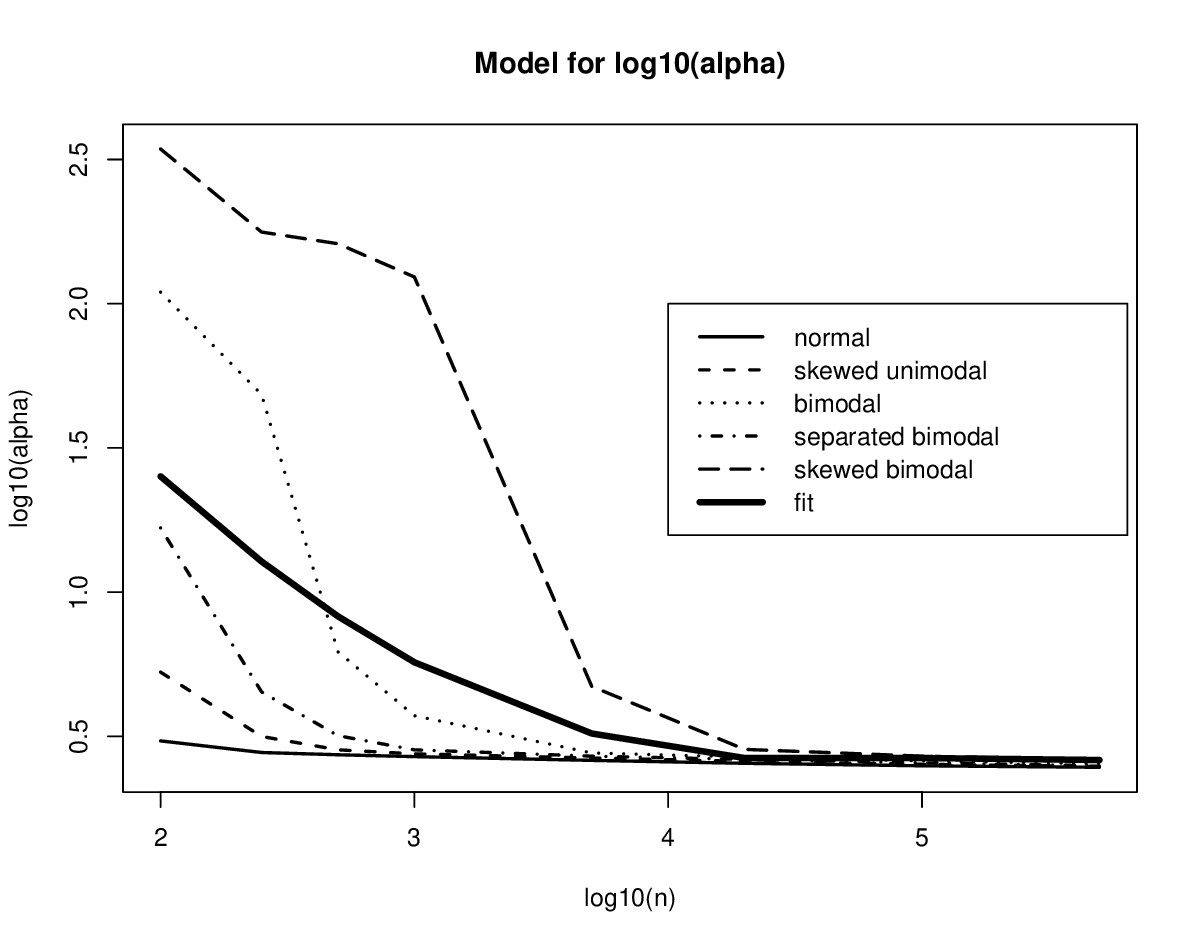,height=280pt}\\
\epsfig{file=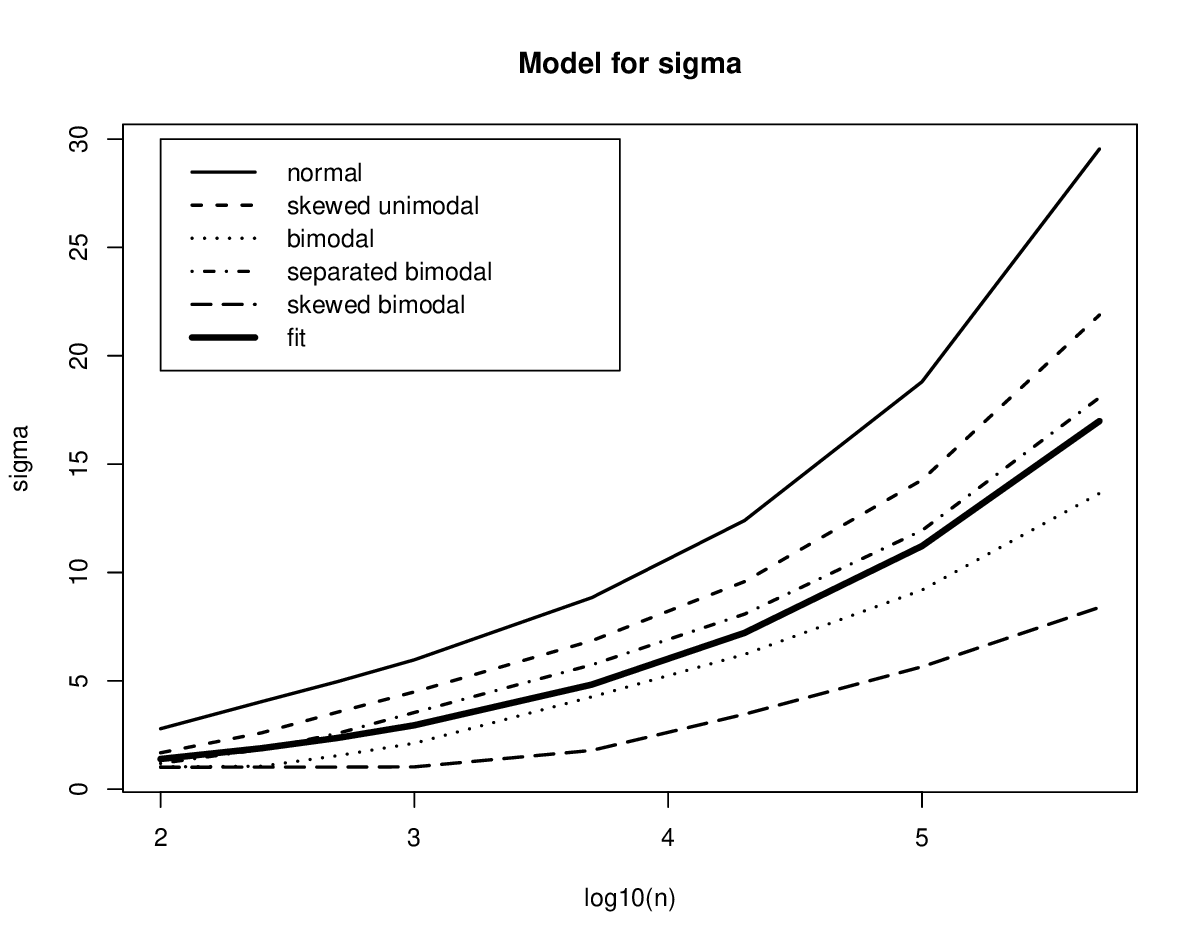,height=280pt}\\
\end{tabular}
\caption{MSE-optimal $\log_{10}(\alpha)$ and $\sigma$ and the model
fits. \label{fig:model}}
\end{center}
\end{figure}

We found a practical purpose model for $\alpha$ and $\sigma$ by
using polynomial regression. Our independent variable was
$\log_{10}(n)$ and our dependent variables were the MSE-optimal
values of $\log_{10}(\alpha)$ and $\log_{10}(\sigma)$ for different
densities. The $\log_{10}$ transformations for $\alpha$ and $\sigma$
stabilize variability. Using a sixth degree polynomial for $\alpha$
and a quadratic for $\sigma$, we arrived at the following models for
$\alpha$ and $\sigma$:

\begin{equation}
\label{eq:model}
\begin{array}{l}
\alpha_{mod}=10^{3.390-1.093\log10(n)+0.025\log10(n)^3-0.00004\log10(n)^6}\\
\sigma_{mod}=10^{-0.58+0.386\log10(n)-0.012\log10(n)^2},\\
\end{array}
\end{equation}
which are appropriate for $100\leq n\leq500000$. The MSE-optimal
values of $\log_{10}(\alpha)$ and $\sigma$ together with the model
fits are shown in Figure~\ref{fig:model}.
In Table~\ref{tab:modchoice} we give the model choices
$\alpha_{mod}$ and $\sigma_{mod}$ for the same sample sizes as in
Table~\ref{tab:MSE_min}.

\begin{table}
\begin{center}
\begin{tabular}{|c|c|c|c|c|c|c|c|c|}
\hline $n$&100&250&500&1000&5000&20000&100000&500000\\
\hline $\alpha_{mod}$&25.20&12.77&8.24&5.71&3.23&2.66&2.66&2.62\\
\hline $\sigma_{mod}$&1.39&1.89&2.37&2.95&4.83&7.21&11.22&16.98\\
\hline
\end{tabular}
\end{center}
\caption{Model choices of $\alpha$ and
$\sigma$.\label{tab:modchoice}}
\end{table}

\section{Simulation study \label{sec:Sims}}

The primary goal of our simulation study was to compare ICV with
ordinary LSCV. However, we will also provide simulation results for
the Sheather-Jones plug-in method.

We considered the four sample sizes $n=100$, 250, 500 and 5000, and
took samples from the target densities listed in
Section~\ref{sec:MSEopt}. For each combination of density and sample
size we did 1000 replications. In all cases the parameters $\alpha$
and $\sigma$ in the selection kernel $L$ were chosen according to
model (\ref{eq:model}).

Let $\hat h_0$ denote the minimizer of $ISE(h)$ for a Gaussian
kernel estimator. For each sample, we computed $\hat h_0$, $\hat
h_{ICV}^*$, $\hat h_{UCV}$ and the Sheather-Jones plug-in bandwidth
$\hat h_{SJPI}$. The definition of $\hat h_{ICV}^*$ is as follows:
\begin{equation}
\label{eq:h.ICV*}
\hat h_{ICV}^*=\min(\hat h_{ICV},\hat h_{OS}),
\end{equation}
where $\hat h_{OS}$ is the oversmoothed bandwidth
of~\citeN{Terrell:OS}. It is arguable that {\it no} data-driven
bandwidth should be larger than $\hat h_{OS}$ since this statistic
estimates an upper bound for {\it all} MISE-optimal bandwidths
(under standard smoothness conditions).

For any random variable $Y$ defined in each replication of our
simulation, we denote the mean, standard deviation and median of $Y$
over all replications (with $n$ and $f$ fixed) by $\widehat{\E}(Y)$,
$\widehat{\SD}(Y)$ and $\widehat{\mbox{Median}}(Y)$. To evaluate the
bandwidth selectors we computed $\widehat{\E}\bigl\{ISE(\hat
h)/ISE(\hat h_0)\bigr\}$ and $\widehat{\mbox{Median}}\bigl\{ISE(\hat
h)/ISE(\hat h_0)\bigr\}$ for $\hat h$ equal to each of $\hat
h_{ICV}^*$, $\hat h_{UCV}$ and $\hat h_{SJPI}$. We also computed the
performance measure $\widehat E\left(\hat h-\hat E(\hat
h_0)\right)^2$, which estimates the MSE of the bandwidth $\hat h$.

\begin{figure}
\begin{center}
\begin{tabular}{cc}
$n=100$&$n=250$\\
\epsfig{file=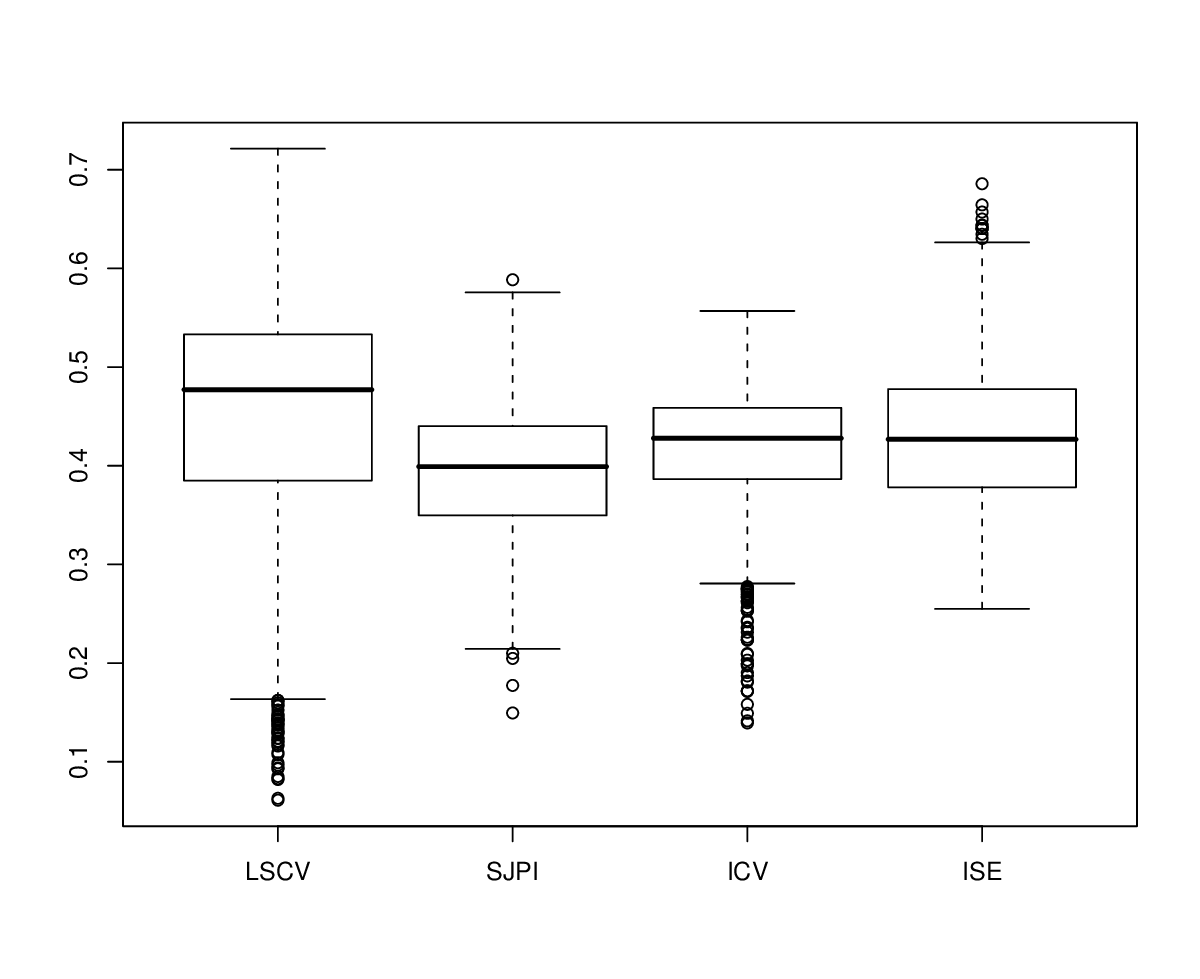,height=180pt}&\epsfig{file=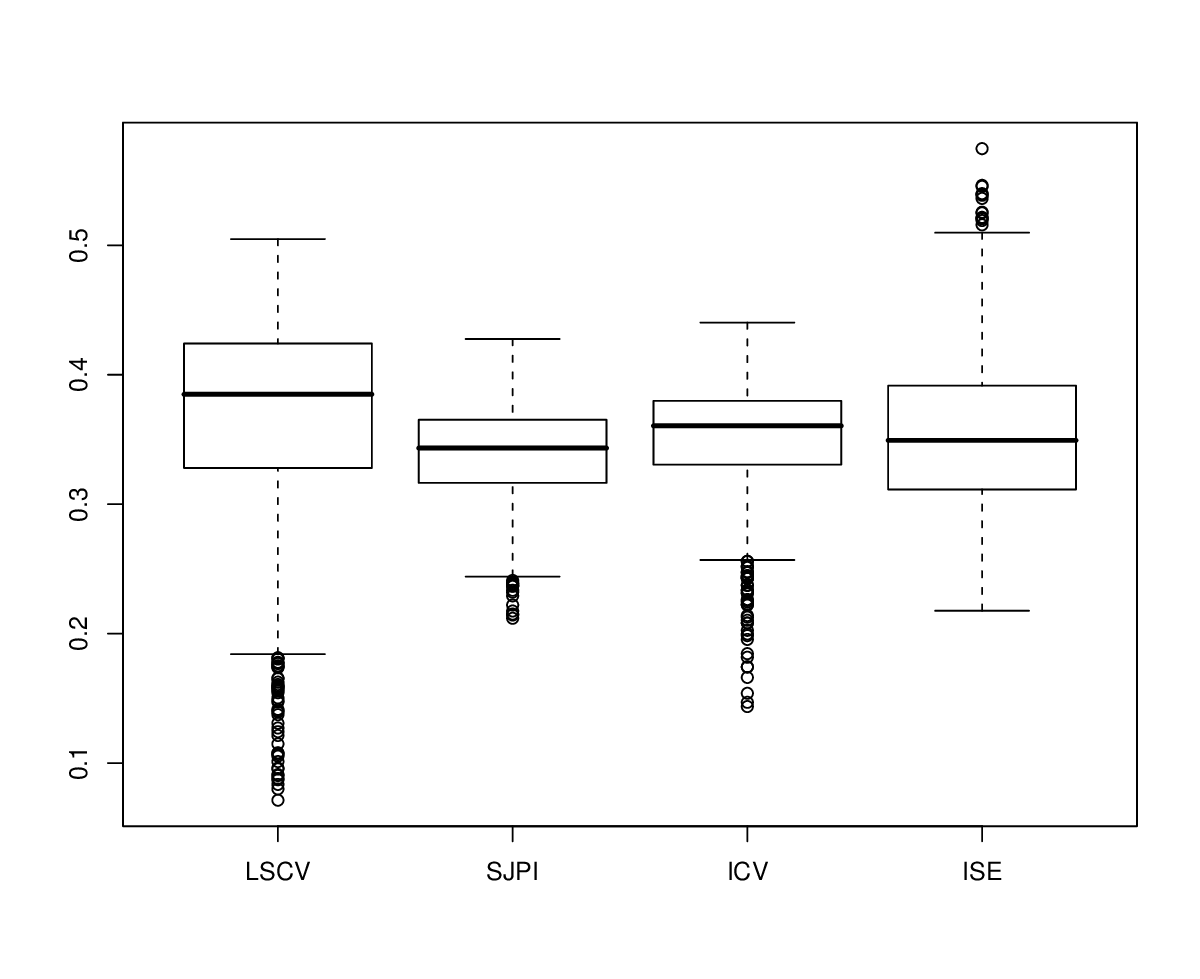,height=180pt}\\
$n=500$&$n=5000$\\
\epsfig{file=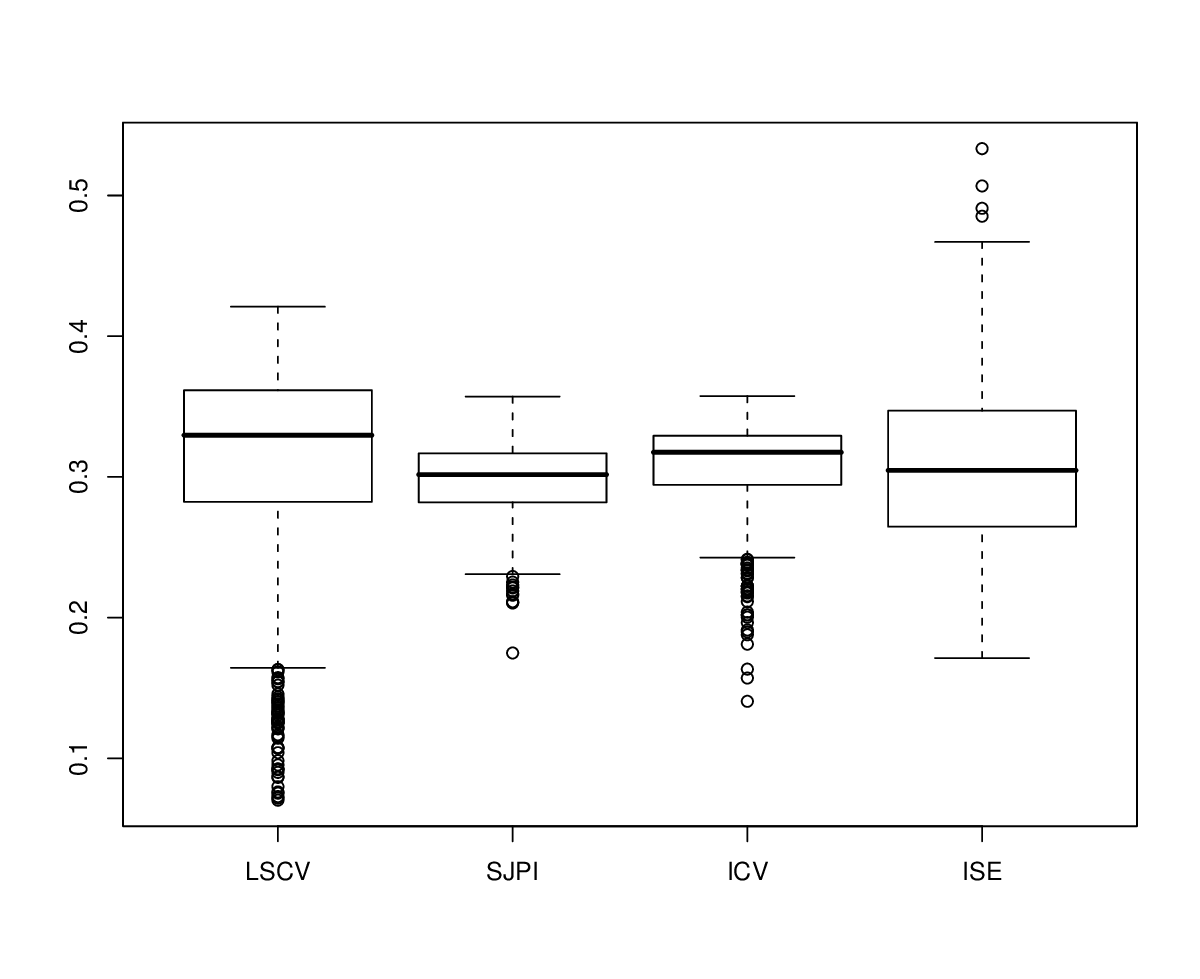,height=180pt}&\epsfig{file=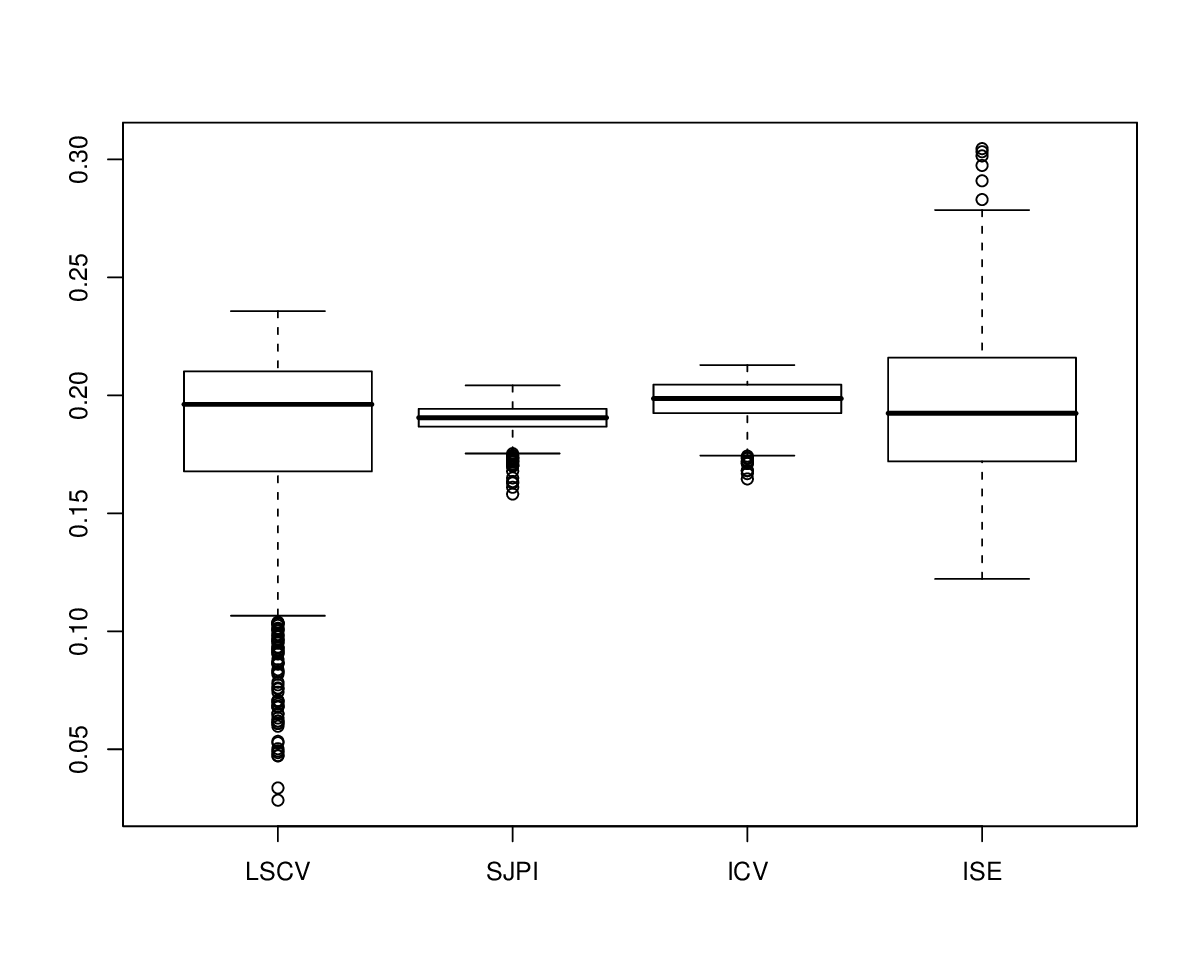,height=180pt}\\
\end{tabular}
\caption{Boxplots for the data-driven bandwidths in case of the {\bf
Normal density}.\label{fig:norm}}
\end{center}
\end{figure}

\begin{figure}
\begin{center}
\begin{tabular}{cc}
$n=100$&$n=250$\\
\epsfig{file=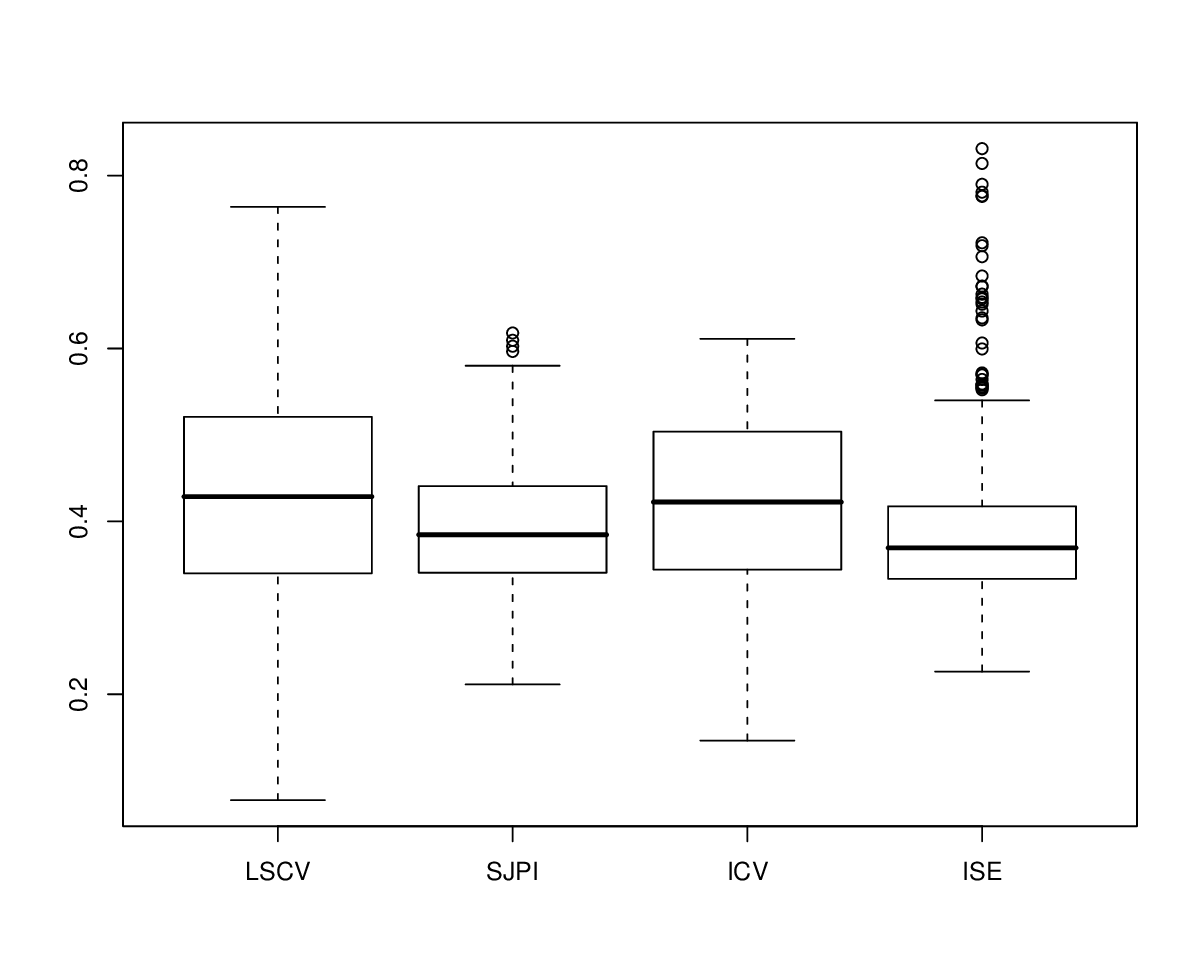,height=180pt}&\epsfig{file=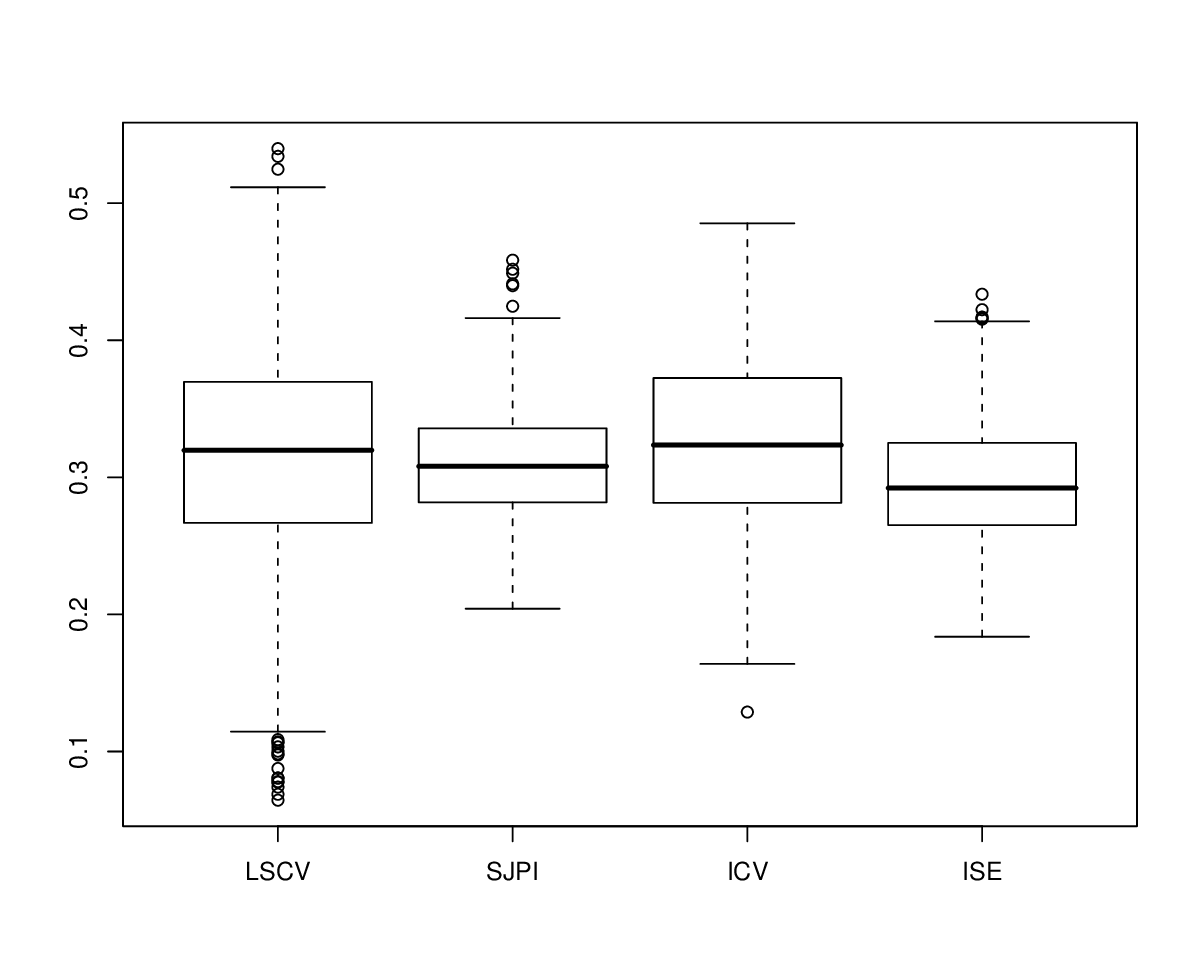,height=180pt}\\
$n=500$&$n=5000$\\
\epsfig{file=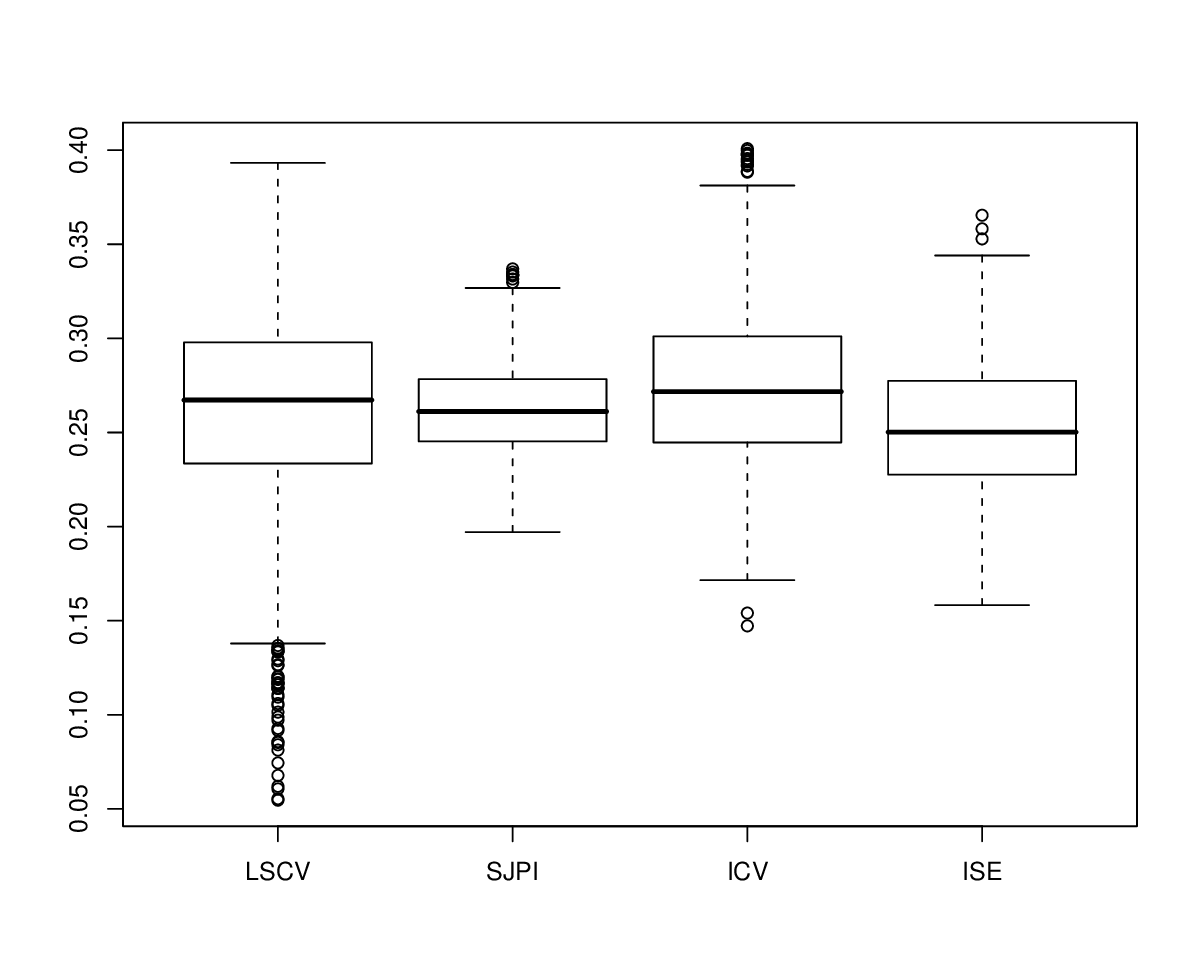,height=180pt}&\epsfig{file=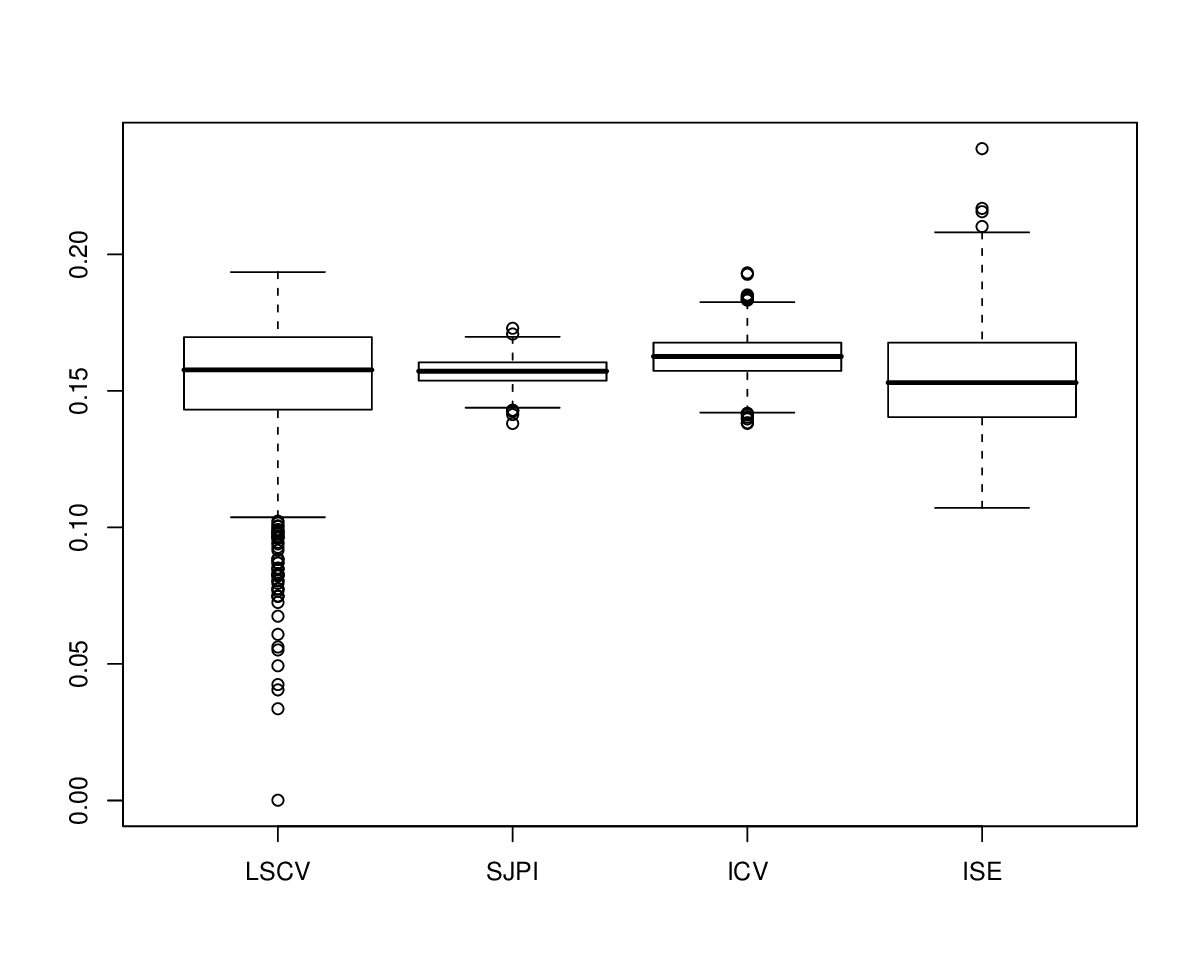,height=180pt}\\
\end{tabular}
\caption{Boxplots for the data-driven bandwidths in case of the {\bf
Bimodal density}. \label{fig:bimod}}
\end{center}
\end{figure}

Our main simulation results for the ``normal'' and ``bimodal''
densities, as defined in Section \ref{sec:MSEopt}, are given in
Tables~\ref{tab:norm} and~\ref{tab:bimod} and Figures~\ref{fig:norm}
and~\ref{fig:bimod}.  Results for the other densities are available
from the authors. Other statistics reported in Tables~\ref{tab:norm}
and~\ref{tab:bimod} are $\widehat{\E}(\hat h)$ and
$\widehat{\SD}(\hat h)$ for each type of bandwidth considered.

\begin{table}
\begin{center}
{\small
\begin{tabular}{|c||c|c|c|c|}
\hline $n$&\textbf{LSCV}&\textbf{SJPI}&\textbf{ICV}&\textbf{ISE}\\
\hline
\hline \multicolumn{5}{|c|}{$\widehat{\E}(\hat{h})$}\\
\hline 100&0.44524596&0.39338747&0.41530230&0.43162318\\
\hline 250&0.36398008&0.33883538&0.34944737&0.35487029\\
\hline 500&0.31094126&0.29803205&0.30864570&0.30806146\\
\hline 5000&0.18359629&0.18992356&0.19768683&0.19526358\\
\hline
\hline \multicolumn{5}{|c|}{$\widehat{\SD}(\hat{h})\cdot10^2$}\\
\hline 100&12.32173263&6.43244579&6.52298637&7.52008697\\
\hline 250&8.35772162&3.71742374&4.44775700&6.27300326\\
\hline 500&7.11168918&2.60300987&3.08015801&5.63495059\\
\hline 5000&3.90077096&0.61900268&0.82041632&3.09277421\\
\hline
\hline \multicolumn{5}{|c|}{$\widehat{\E}(\hat{h}-\widehat{\E}(\hat{h}_0))^2\cdot10^4$}\\
\hline 100&153.52907115&55.95467615&45.17051435&\\
\hline 250&70.61154173&16.37660421&20.05684094&\\
\hline 500&50.60847941&7.77477660&9.48129936&\\
\hline 5000&16.56205491&0.66793916&0.73113122&\\
\hline
\hline \multicolumn{5}{|c|}{$\widehat{\E}\bigl(\ISE(\hat{h})/\ISE(\hat h_0)\bigr)$}\\
\hline 100&2.46997542&1.90795915&1.72178966&\\
\hline 250&1.91593730&1.50563016&1.47567596&\\
\hline 500&1.75806058&1.37734003&1.36096679&\\
\hline 5000&1.41316047&1.11460567&1.10313807&\\
\hline
\hline \multicolumn{5}{|c|}{$\widehat{\mbox{Median}}\bigl(\ISE(\hat{h})/\ISE(\hat h_0)\bigr)$}\\
\hline 100&1.31108630&1.15695876&1.11233574&\\
\hline 250&1.21715835&1.10408948&1.09365380&\\
\hline 500&1.21396609&1.10306404&1.09608944&\\
\hline 5000&1.10907960&1.04471055&1.05183075&\\
\hline
\end{tabular}
\caption{Simulation results for the {\bf Gaussian
density}.\label{tab:norm} }}
\end{center}
\end{table}

\begin{table}
\begin{center}
{\small
\begin{tabular}{|c||c|c|c|c|}
\hline $n$&\textbf{LSCV}&\textbf{SJPI}&\textbf{ICV}&\textbf{ISE}\\
\hline
\hline \multicolumn{5}{|c|}{$\widehat{\E}(\hat{h})$}\\
\hline 100&0.42908686&0.39453431&0.41955286&0.38237337\\
\hline 250&0.31360942&0.31160054&0.32846189&0.29715278\\
\hline 500&0.25927533&0.26238646&0.27450416&0.25320682\\
\hline 5000&0.15262210&0.15706804&0.16255246&0.15478049\\
\hline
\hline \multicolumn{5}{|c|}{$\widehat{\SD}(\hat{h})\cdot10^2$}\\
\hline 100&13.56532316&7.44425312&9.56680379&7.60899932\\
\hline 250&8.46734473&4.18778288&6.50918853&4.29431763\\
\hline 500&5.70587208&2.44443305&4.20078840&3.55982408\\
\hline 5000&2.46293965&0.47951752&0.81457083&1.96503777\\
\hline
\hline \multicolumn{5}{|c|}{$\widehat{\E}(\hat{h}-\widehat{\E}(\hat{h}_0))^2\cdot10^4$}\\
\hline 100&205.65547766&56.84037076&105.25535253&\\
\hline 250&74.33244070&19.60736507&52.12977298&\\
\hline 500&32.89268754&6.81193546&22.16474597&\\
\hline 5000&6.10659189&0.28203637&1.26689717&\\
\hline
\hline \multicolumn{5}{|c|}{$\widehat{\E}\bigl(\ISE(\hat{h})/\ISE(\hat h_0)\bigr)$}\\
\hline 100&1.69951929&1.32733595&1.36143018&\\
\hline 250&1.51599857&1.20914143&1.28743335&\\
\hline 500&1.41670996&1.15070890&1.19168891&\\
\hline 5000&2.06430484&1.06839987&1.07675906&\\
\hline
\hline \multicolumn{5}{|c|}{$\widehat{\mbox{Median}}\bigl(\ISE(\hat{h})/\ISE(\hat h_0)\bigr)$}\\
\hline 100&1.20951575&1.08744161&1.13356965&\\
\hline 250&1.16087896&1.08338970&1.12699702&\\
\hline 500&1.12243694&1.06072702&1.09421867&\\
\hline 5000&1.05825025&1.03067963&1.03649944&\\
\hline
\end{tabular}
\caption{Simulation results for the {\bf Bimodal
density}.\label{tab:bimod}}}
\end{center}
\end{table}

The reduced variability of the ICV bandwidth is evident in our
study. The ratio $\widehat{\mbox{SD}}(\hat
h_{ICV}^*)/\widehat{\mbox{SD}}(\hat h_{UCV})$ ranged between 0.9713
and 0.2103 in the twenty settings considered. However, the variances
of the ICV bandwidths were always higher compared to the
Sheather-Jones plug-in bandwidths. It is worth noting that the ratio
of sample standard deviations of the ICV and LSCV bandwidths
decreases as the sample size $n$ increases.

The mean squared distance $\widehat {\E}\left(\hat h-\widehat
{\E}(\hat h_0)\right)^2$ was smaller for the ICV method than for the
LSCV method in all but two cases corresponding to the Skewed Bimodal
density, $n=250$ and 500. Plug-in always had a smaller value of
$\widehat {\E}\left(\hat h-\widehat{\E}(\hat h_0)\right)^2$ than did
ICV.

The most important observation is that the values of
$\widehat{\E}\bigl(ISE(\hat{h})/ISE(\hat h_0)\bigr)$ were smaller
for ICV than for LSCV for all combinations of densities and sample
sizes. The values of
$\widehat{\mbox{Median}}\bigl(ISE(\hat{h})/ISE(\hat h_0)\bigr)$ were
smaller for ICV than for LSCV in all but one case, which corresponds
to the Skewed Bimodal density at $n=250$ when
$\widehat{\mbox{Median}}\bigl(ISE(\hat{h}_{ICV})/ISE(\hat
h_0)\bigr)$ was 1.0013 times greater than
$\widehat{\mbox{Median}}\bigl(ISE(\hat{h}_{UCV})/ISE(\hat
h_0)\bigr)$.

\begin{figure}
\begin{center}
\epsfig{file=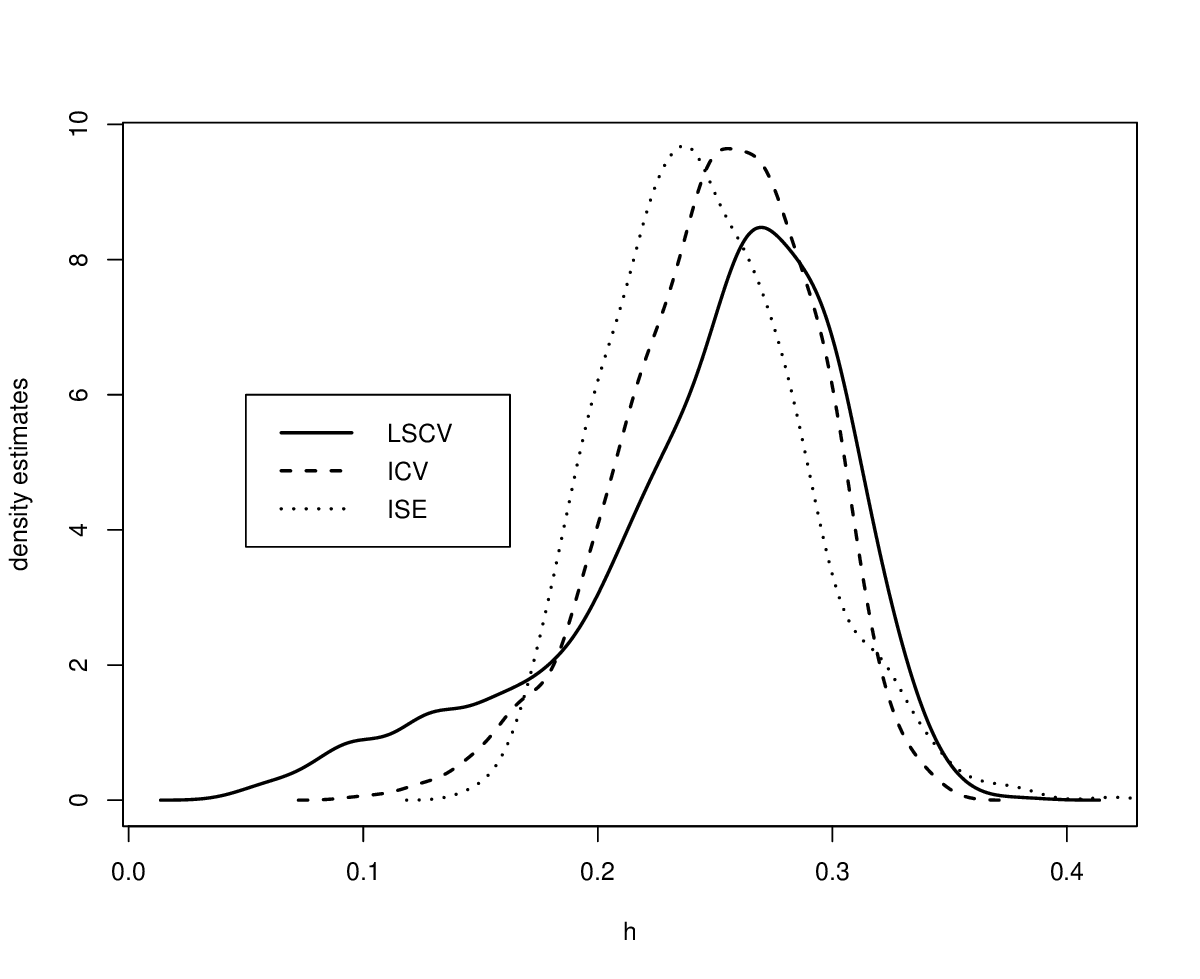,height=200pt} \caption{Kernel
density estimates for random bandwidths from the simulation with the
Skewed Unimodal density and $n=250$. \label{fig:banddistr}}
\end{center}
\end{figure}

Despite the fact that the LSCV bandwidth is asymptotically normally
distributed (see~\citeN{H&M:Extent}), its distribution in finite
samples tends to be skewed to the left. In our simulations we have
noticed that the distribution of the ICV bandwidth is less skewed
than that of the LSCV bandwidth.  A typical case is illustrated in
Figure~\ref{fig:banddistr}, where kernel density estimates for the
two data-driven bandwidths are plotted from the simulation with the
Skewed Unimodal density at $n=250$. Also plotted is a density
estimate for the ISE-optimal bandwidths. Note that the ICV density
is more concentrated near the middle of the ISE-optimal distribution
than the density estimate for LSCV.

\begin{figure}
\begin{center}
\begin{tabular}{cc}
\textbf{(a)}&\textbf{(b)}\\
\epsfig{file=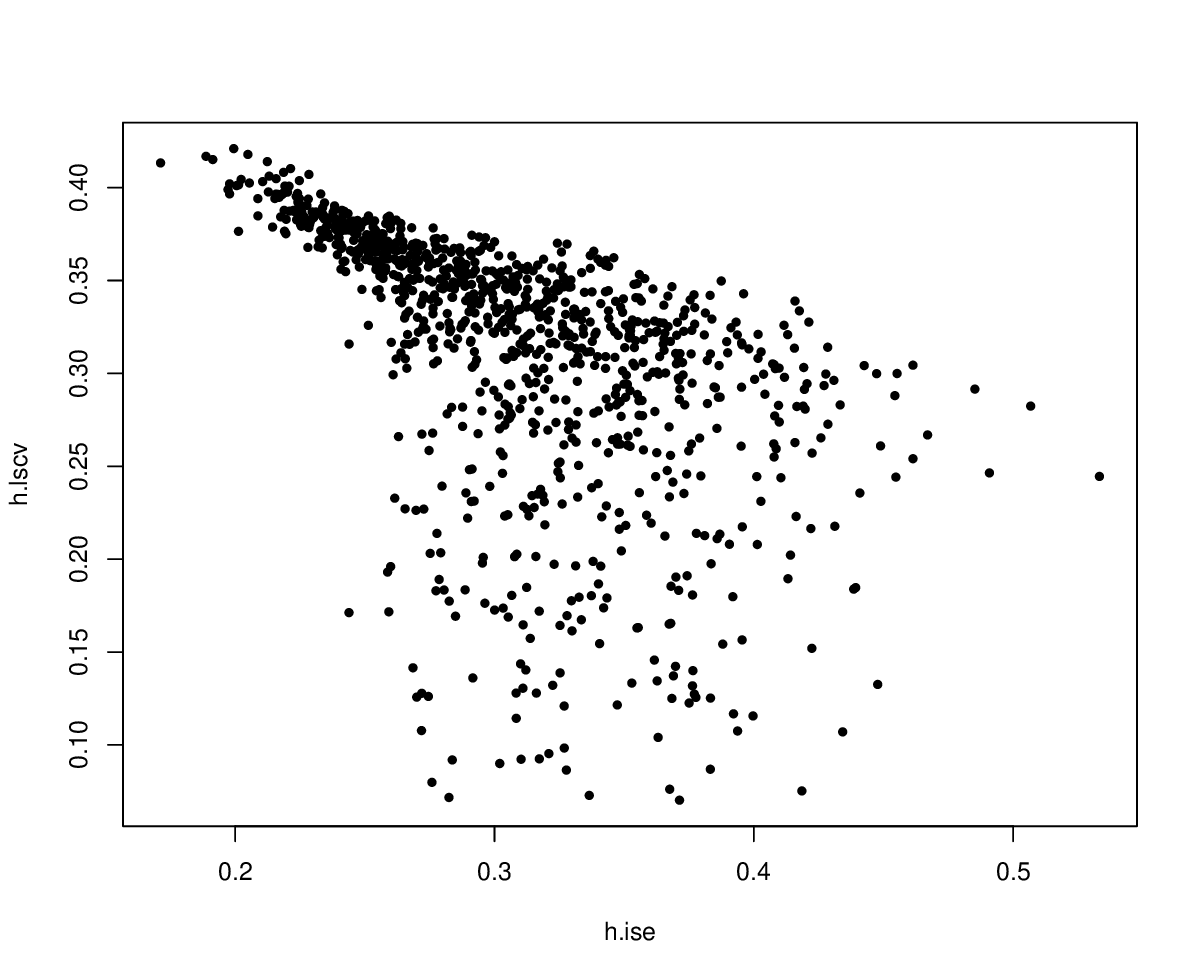,height=175pt}&\epsfig{file=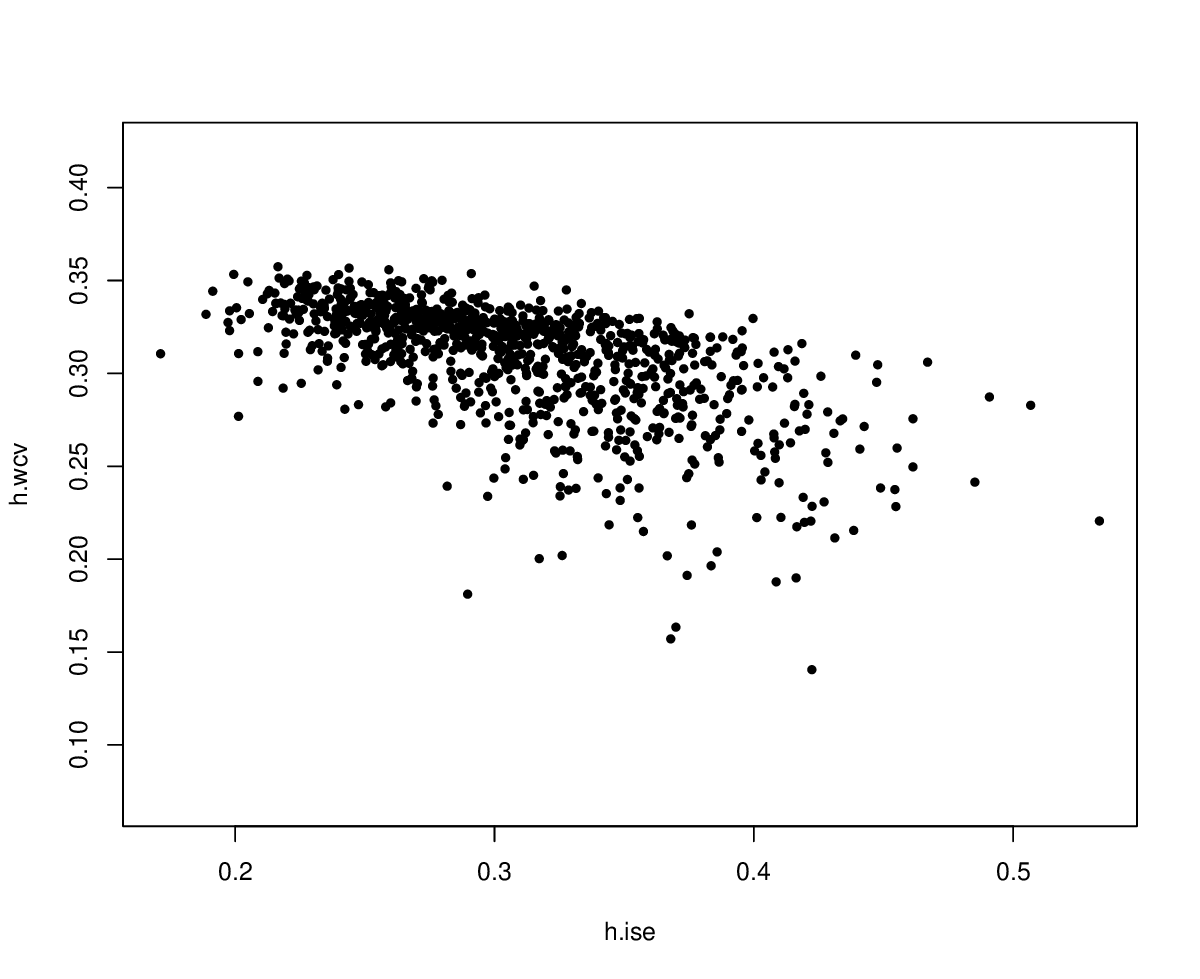,height=175pt}\\
\end{tabular}
\caption{Scatterplots of $\hat h$ vs. $\hat h_0$ for the case of a
Gaussian density and $n=500$, with $\hat h$ corresponding to the
\textbf{(a)} LSCV and \textbf{(b)} ICV bandwidths.
\label{fig:scatterband}}
\end{center}
\end{figure}

Figure~\ref{fig:scatterband} provides scatterplots of the bandwidths
$\hat h_{UCV}$ and $\hat h_{ICV}$ versus $\hat h_0$ in the case of
the Gaussian density and $n=500$. The sample correlation
coefficients were -0.52 and -0.60 for LSCV and ICV, respectively.
The fact that these correlations are negative is a well-established
phenomenon; see, for example, \citeN{HJ}. Note that the ICV
bandwidths cluster more tightly about the MISE minimizer
$h_0=0.315$.

\begin{figure}[h]
\begin{center}
\begin{tabular}{cc}
{\bf (a)}&{\bf (b)}\\
\epsfig{file=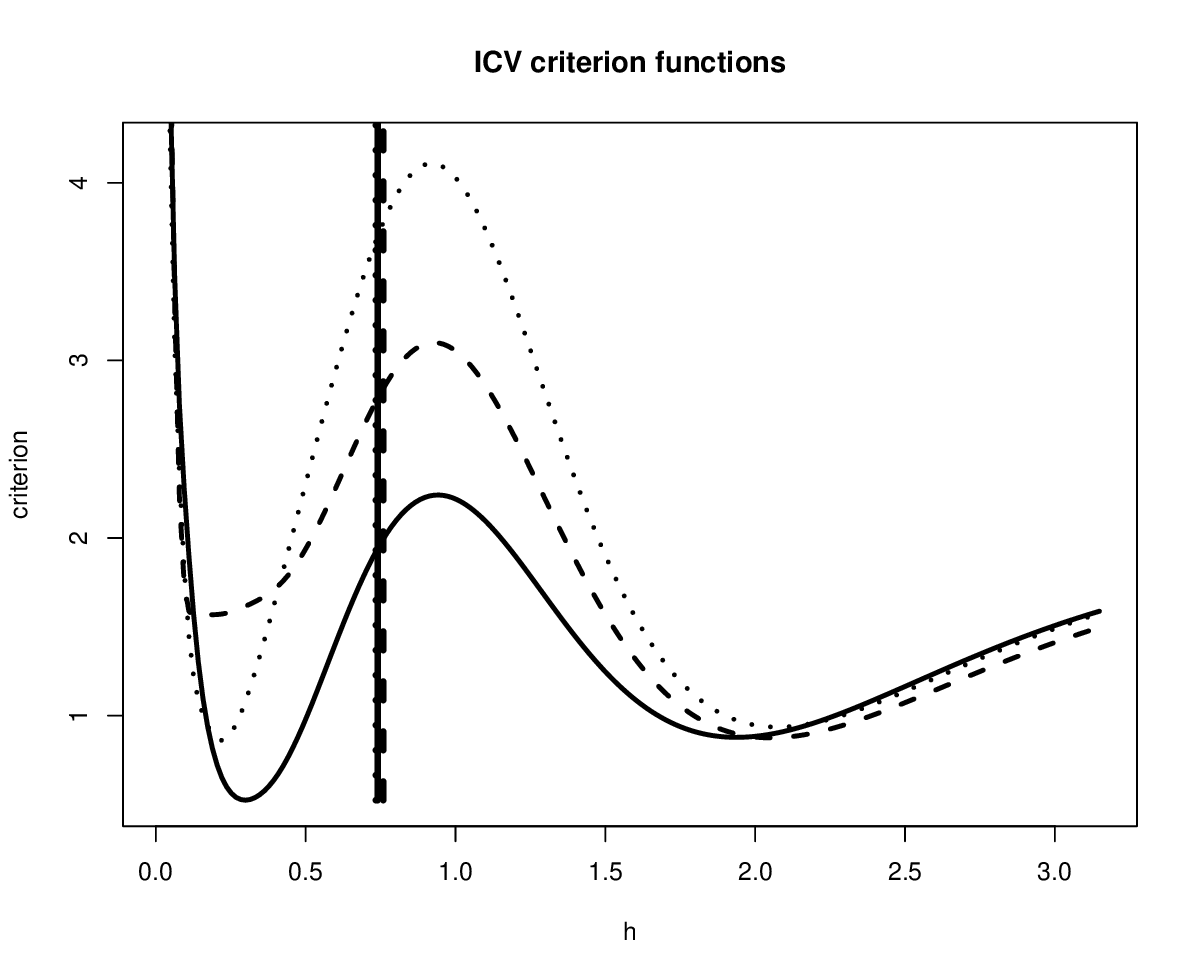,height=175pt}&\epsfig{file=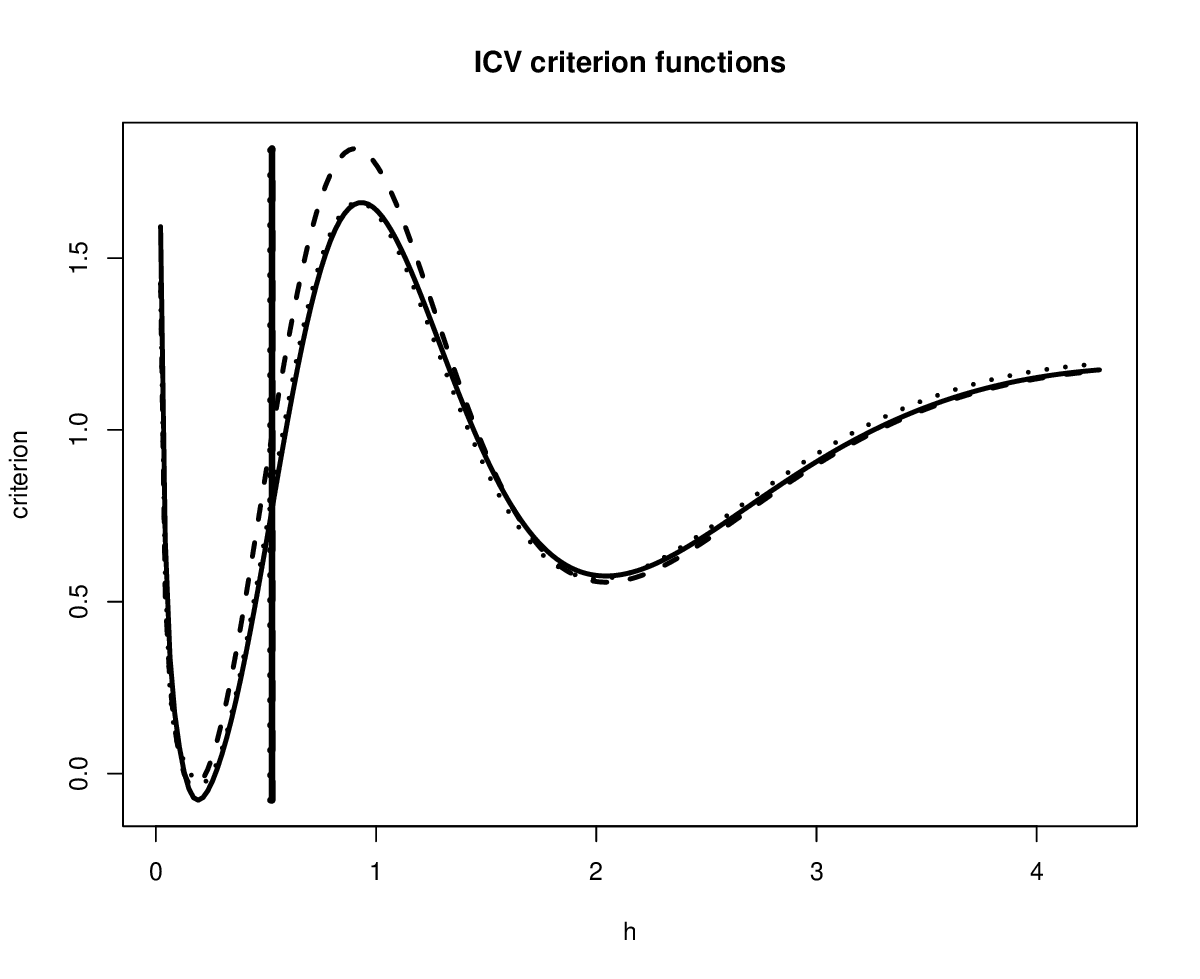,height=175pt}\\
\end{tabular}
\caption{Three ICV criterion functions in case of the Separated
Bimodal density at {\bf (a)} $n=100$ and {\bf(b)}
$n=500$.\label{fig:criterion}}
\end{center}
\end{figure}

A problem we have noticed with the ICV method is that its criterion
function can have two local minima when the sample size is moderate
and the density has two modes. The following example illustrates the
problem. In Figure~\ref{fig:criterion}(a) we have plotted three ICV
curves for the case of the Separated Bimodal density and $n=100$.
The minimizers of the solid, dashed and dotted lines occur at the
$h$-values 0.2991, 2.0467 and 0.2204, respectively. For comparison,
the corresponding bandwidths chosen by the Sheather-Jones plug-in
method are 0.3240, 0.2508 and 0.2467. The value of $h=2.0467$ which
minimizes the dashed ICV curve is obviously too large. The local
minimum at 0.1295 would yield a much more reasonable estimate. The
problem of choosing too large a bandwidth from the second local
minimum is mitigated by using the rule~\eqref{eq:h.ICV*}. Indeed,
the oversmoothed bandwidths for the three samples are shown by the
vertical lines in Figure~\ref{fig:criterion} and were 0.7404, 0.7580
and 0.7341. Note that the problem with the ICV curve having two
local minima of approximately the same value quickly goes away as
the sample size increases. This is illustrated in
Figure~\ref{fig:criterion}(b), where we have plotted three criterion
curves for the Separated Bimodal case with $n=500$. Thus, the
selection rule $\hat h_{ICV}^*$ given by~\eqref{eq:h.ICV*} rather
than just $\hat h_{ICV}$ appears to be useful mostly for small and
moderate sample sizes.

\section{Real data examples\label{sec:Data}}

In this section we show how the ICV method works on two real data
sets. The purpose of the first example is to compare the performance
of the ICV, LSCV, and Sheather-Jones plug-in methods. The second
example illustrates the benefit of using ICV locally.

\subsection{PGA data}
In this example the data are the average numbers of putts per round
played, for the top 175 players on the 1980 and 2001 PGA golf tours.
The question of interest is whether there has been any improvement
from 1980 to  2001. This data set has already been analyzed
by~\citeN{Sheather:Density} in the context of comparing the
performances of LSCV and Sheather-Jones plug-in.

\begin{figure}[h]
\begin{center}
\epsfig{file=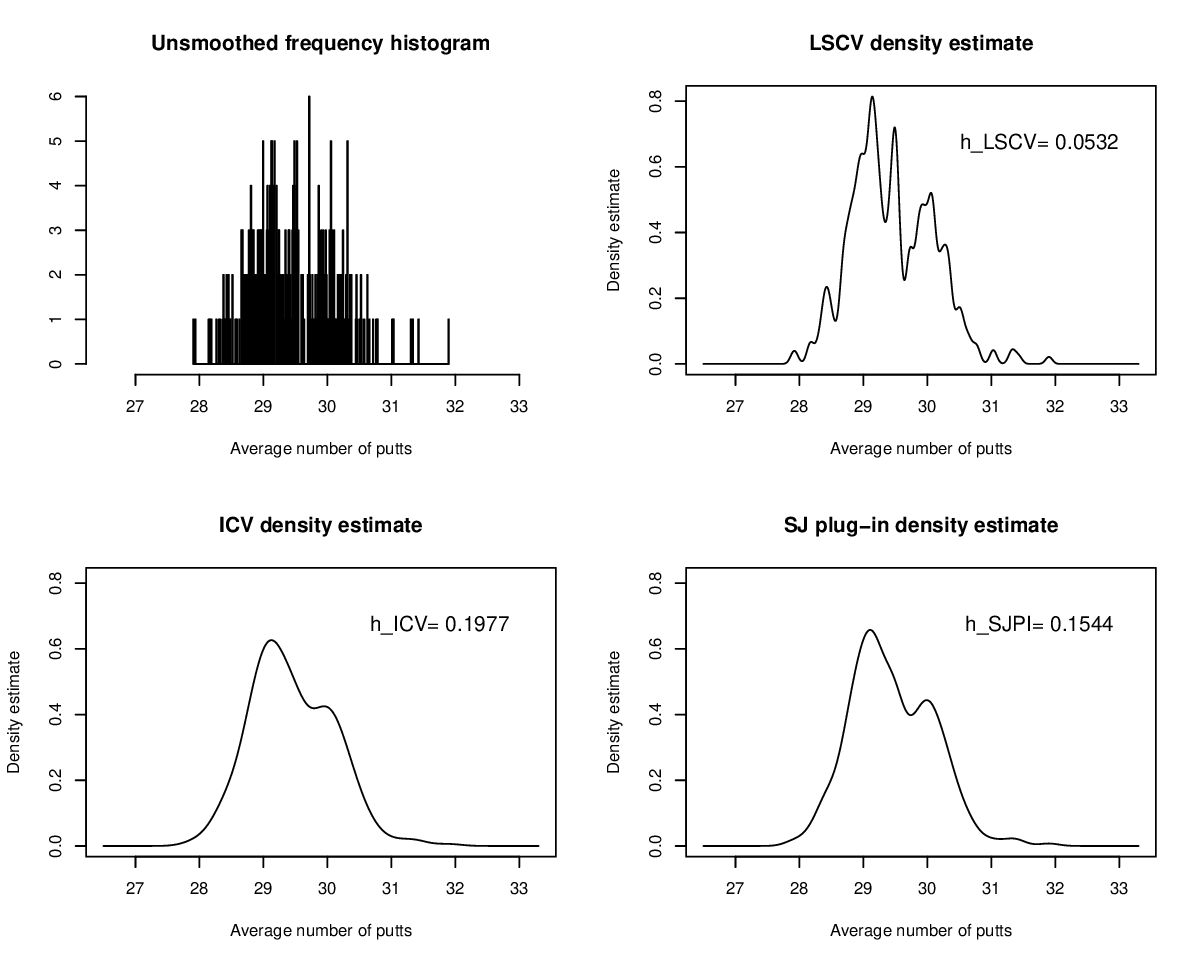,height=365pt} \caption{Unsmoothed frequency
histogram and kernel density estimates for average numbers of putts
per round from 1980 and 2001 combined. \label{fig:KDE_PGA}}
\end{center}
\end{figure}

In Figure~\ref{fig:KDE_PGA} we have plotted an unsmoothed frequency
histogram and the LSCV, ICV and Sheather-Jones plug-in density
estimates for a combined data set of 1980 and 2001 putting averages.
The class interval size in the unsmoothed histogram was chosen to be
0.01, which corresponds to the accuracy to which the data have been
reported. There is a clear indication of two modes in the histogram.

The estimate based on the LSCV bandwidth is apparently
undersmoothed. The ICV and plug-in estimates look similar and have
two modes, which agrees with evidence from the unsmoothed histogram
and seems reasonable since the data were taken from two populations.

\begin{figure}[h]
\begin{center}
\epsfig{file=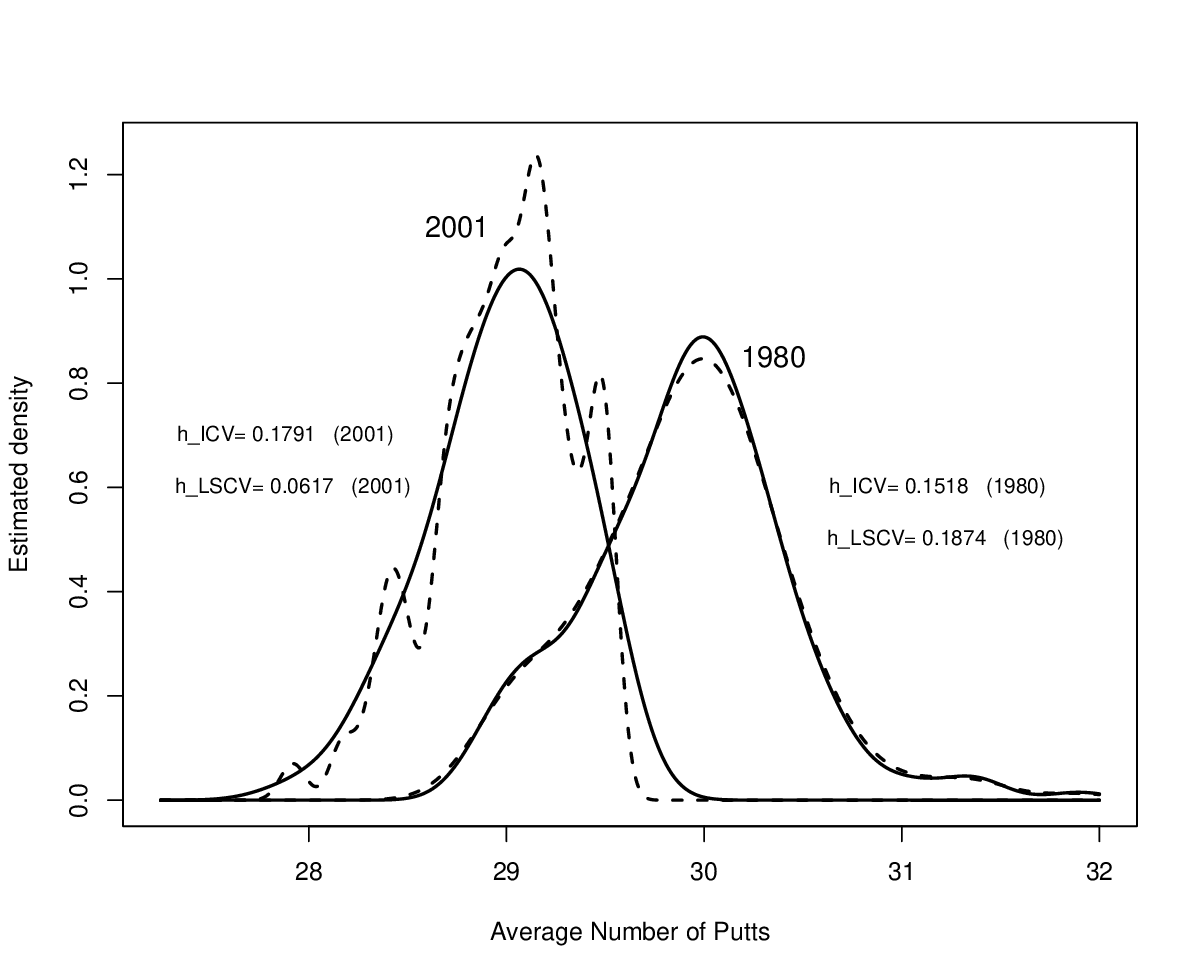,height=350pt} \caption{Kernel density
estimates based on LSCV (dashed curve) and ICV (solid curve)
produced separately for the data from 1980 and
2001.\label{fig:KDE_PGA_sep}}
\end{center}
\end{figure}

In Figure~\ref{fig:KDE_PGA_sep} we have plotted kernel density
estimates separately for the years 1980 and 2001. ICV seems to
produce a reasonable estimate in both years, whereas LSCV yields a
very wiggly and apparently undersmoothed estimate in 2001.

\subsection{Local ICV example}

Local cross-validation methods for density estimation, independently
proposed by~\citeN{Hall&Schucany:LocCV}
and~\citeN{Mielniczuk:LocCV}, consist in performing LSCV at each
value of the argument $x$ using a fraction of the data that are
close to $x$. Allowing the bandwidth to depend on $x$ is desirable
when the smoothness of the underlying density changes sufficiently
with $x$.

The local ICV method was introduced
in~\citeN{SavchukHartSheather:ICV}. It is different from the local
LSCV method in that it uses ICV rather than LSCV for the local
bandwidth selection. Another difference is that local ICV uses the
first local minimizer of the local criterion function as opposed to
the global minimizer of local LSCV.

The local ICV criterion function is defined as
\[
ICV(x,b,w)=\frac{1}{w}\int \phi\left(\frac{x-u}{w}\right)\hat
f_b(u)^2\,du - \frac{2}{nw}\sum_{i=1}^n
\phi\left(\frac{x-X_i}{w}\right)\hat f_{b,-i}(X_i),
\]
where function $\hat f_b$ is the kernel density estimate based on a
selection kernel $L$ with a smoothing parameter $b$. The quantity
$w$ defines the extent to which the cross-validation is local, with
a large choice of $w$ corresponding to global ICV. Let $\hat b(x)$
be the first local minimum of the local ICV curve for the fixed
value of $x$. Then the corresponding bandwidth of a $\phi-$kernel
estimator is defined as $\hat h(x)=C\hat b(x)$, where $C$ is
computed as in~\eqref{eq:C}. Local ICV outperformed the local LSCV
method in a simulated data example in the article
of~\citeN{SavchukHartSheather:ICV}. In this paper we show how local
ICV and LSCV perform in a real data example.

We analyze the data of size $n=517$ on the Drought Code (DC) of the
Canadian Forest Fire Weather index (FWI) system. DC is one of the
explanatory variables which can be used to predict the burned area
of a forest in the Forest Fires data set. This data can be
downloaded from the website
\url{http://archive.ics.uci.edu/ml/datasets/Forest+Fires}. The data
were collected and analyzed by~\citeN{Forest}.

\begin{figure}
\begin{center}
\begin{tabular}{cc}
{\bf(a)}&{\bf(b)}\\
\epsfig{file=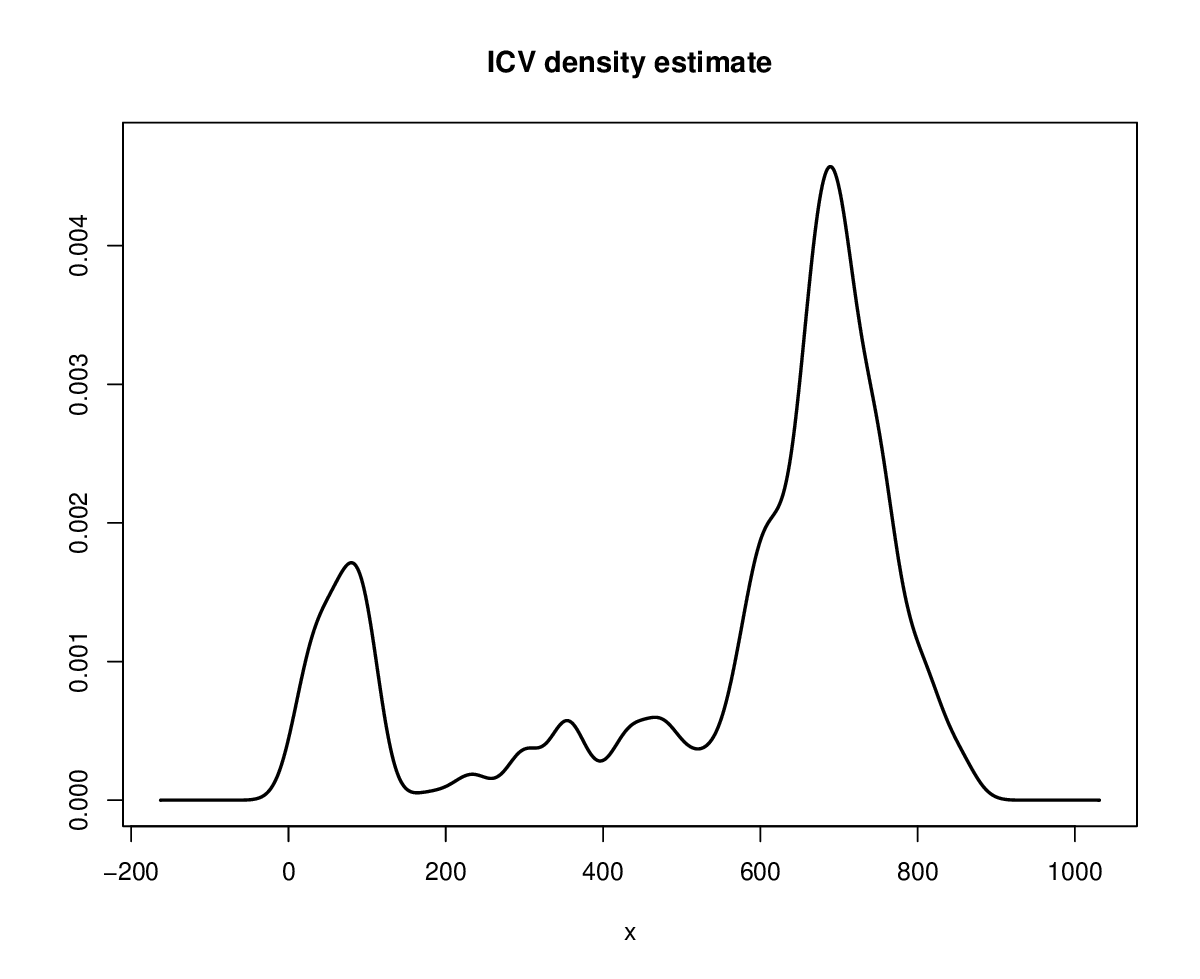,height=170pt}&\epsfig{file=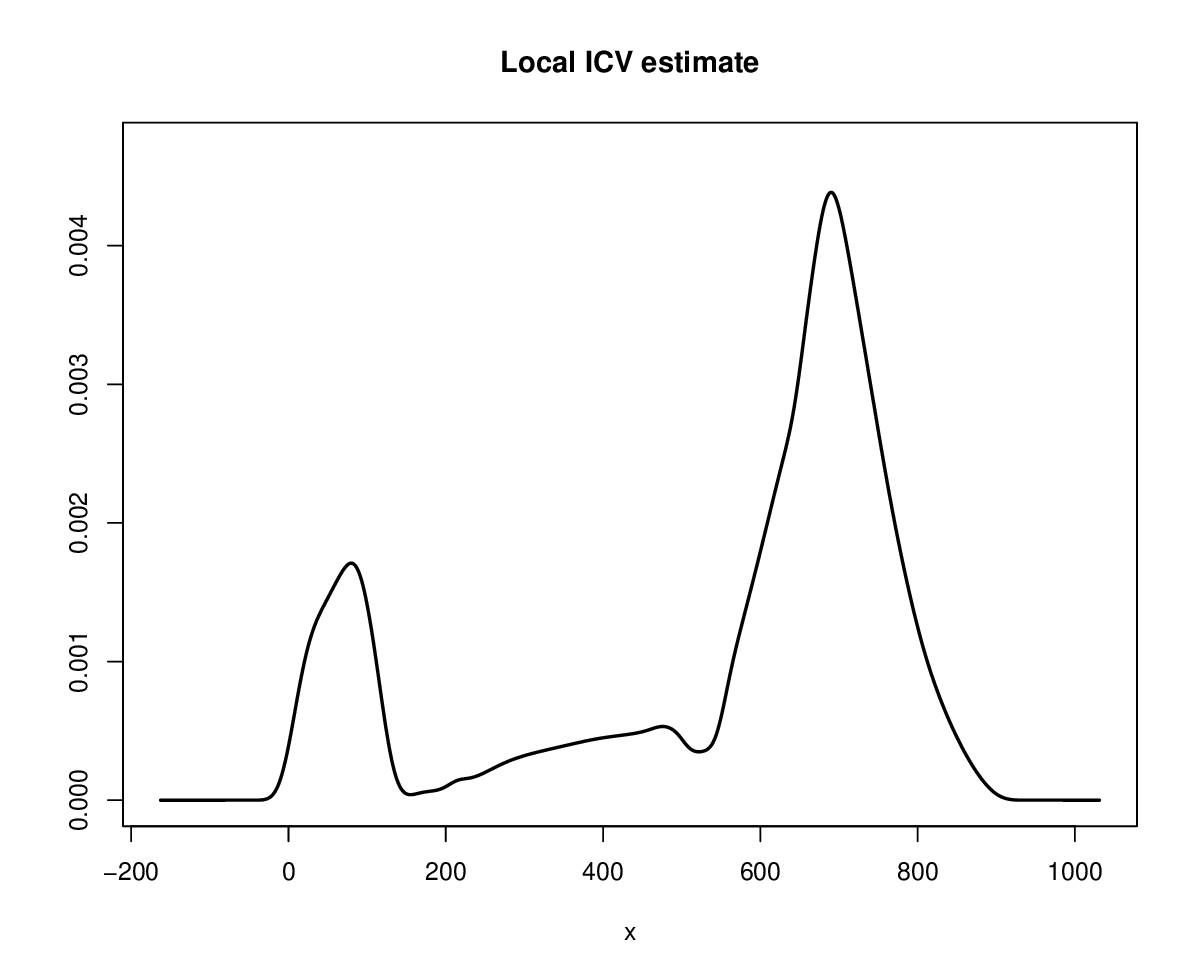,height=170pt}\\
\end{tabular}
\caption{Density estimates for the DC data set with {\bf(a)} being
the global ICV density estimate and {\bf (b)} corresponding to the
local ICV estimate. \label{fig:DC}}
\end{center}
\end{figure}

We computed the LSCV, ICV and Sheather-Jones plug-in bandwidths for
the DC data. The LSCV method failed by yielding $\hat h_{UCV}=0$.
ICV and Sheather-Jones plug-in bandwidths were very close and
produced similar density estimates. Figure~\ref{fig:DC}~{\bf (a)}
gives the ICV density estimate. It shows two major modes connected
with a wiggly curve, which indicates that varying the bandwidth with
$x$ may yield a smoother estimate of the underlying density.

Local ICV and LSCV have been applied to the DC data. We used $w=40$
for both methods and the selection kernel with $\alpha=6$ and
$\sigma=6$ for local ICV. This $(\alpha,\sigma)$ choice performed
quite well for unimodal densities in our simulation studies on
global ICV, and hence seems to be reasonable for local bandwidth
selection since locally the density should have relatively few
features. Let $x_{(i)}$, $i=1,\ldots,n$, denote the $i$th member of
the ordered sequence of observations. The local ICV and LSCV
bandwidth were found for 50 evenly spaced points in the interval
$x_{(1)}-0.2(x_{(n)}-x_{(1)})\leq x\leq
x_{(n)}+0.2(x_{(n)}-x_{(1)})$. It turns out that in 45 out of 50
cases the local LSCV curve tends to $-\infty$ as $h\ra 0$, which
implies that the local LSCV estimate can not be computed. All 50
local ICV bandwidths were positive. We found a smooth function $\hat
h(x)$ by interpolating at other values of $x$ via a spline. The
corresponding local ICV estimate, given in
Figure~\ref{fig:DC}{\bf(b)}, shows a smoother density estimate.

\section{Summary\label{sec:Summary}}

Indirect cross-validation is a method of bandwidth selection in the
univariate kernel density estimation context. The method first
selects the bandwidth of an $L-$kernel estimator by least squares
cross-validation, and then rescales this bandwidth so that it is
appropriate for use in a Gaussian kernel density estimator.
Selection kernels $L$ have the form
$(1+\alpha)\phi(u)-\alpha\phi(u/\sigma)/\sigma$, where
$\alpha\geq0$, $\sigma>0$ and $\phi$ is the Gaussian kernel. Optimal
kernels from this class yield bandwidths with relative error that
converges to 0 at a rate of $n^{-1/4}$, which is a substantial
improvement over the $n^{-1/10}$ rate of LSCV.

A practical purpose model for the selection kernel parameters,
$\alpha$ and $\sigma$, has been developed. The model was built by
performing polynomial regression on the MSE-optimal values of
$\log_{10}(\alpha)$ and $\log_{10}(\sigma)$ at different sample
sizes for five target densities. Use of this model makes the ICV
method completely automatic.

An extensive simulation study showed that in finite samples ICV is
more stable than LSCV. Although both ICV and LSCV bandwidths are
asymptotically normal, the distribution of the ICV bandwidth for
finite $n$ is usually more symmetric and better concentrated in the
middle of the density for ISE-optimal bandwidths. Using an
oversmoothed bandwidth as an upper bound for the bandwidth search
interval reduces the bias of the method and prevents selecting an
impractically large value of $h$ when the criterion curves exhibit
multiple local minima.

The ICV method performs well in real data examples. ICV applied
locally yields density estimates which are more smooth than
estimates based on a single bandwidth. Often, local ICV estimates
may be found when the local LSCV estimates do not exist.

\bibliographystyle{chicago}

\begin{thebibliography}{}

\bibitem[\protect\citeauthoryear{Ahmad and Ran}{Ahmad and Ran}{2004}]{Ahmad}
Ahmad, I.~A. and I.~S. Ran (2004).
\newblock Kernel contrasts: a data-based method of choosing smoothing
  parameters in nonparametric density estimation.
\newblock {\em J. Nonparametr. Stat.\/}~{\em 16\/}(5), 671--707.

\bibitem[\protect\citeauthoryear{Bowman}{Bowman}{1984}]{Bowman:LSCV}
Bowman, A.~W. (1984).
\newblock An alternative method of cross-validation for the smoothing of
  density estimates.
\newblock {\em Biometrika\/}~{\em 71\/}(2), 353--360.

\bibitem[\protect\citeauthoryear{Cao and Vilar~Fernandez}{Cao and
  Vilar~Fernandez}{1993}]{Cao:Dependence}
Cao, R., Q. d. R.~A. and J.~Vilar~Fernandez (1993).
\newblock Bandwidth selection in nonparametric density estimation under
  dependence : a simulation study.
\newblock {\em Computational Statistics\/}~{\em 8}, 313-- 332.

\bibitem[\protect\citeauthoryear{Chiu}{Chiu}{1991}]{Chiu:CV}
Chiu, S.-T. (1991).
\newblock Bandwidth selection for kernel density estimation.
\newblock {\em Ann. Statist.\/}~{\em 19\/}(4), 1883--1905.

\bibitem[\protect\citeauthoryear{Cortez and Morais}{Cortez and
  Morais}{2007}]{Forest}
Cortez, P. and A.~Morais (2007).
\newblock A data mining approach to predict forest fires using meteorological
  data.
\newblock {\em in J. Neves, M. F. Santos and J. Machado Eds., New Trends in
  Artificial Intelligence, Proceedings of the 13th EPIA 2007 - Portuguese
  Conference on Artificial Intelligence, December, Guimaraes, Portugal\/},
  512--523.

\bibitem[\protect\citeauthoryear{Feluch and Koronacki}{Feluch and
  Koronacki}{1992}]{Feluch:spacing}
Feluch, W. and J.~Koronacki (1992).
\newblock A note on modified cross-validation in density estimation.
\newblock {\em Comput. Statist. Data Anal.\/}~{\em 13\/}(2), 143--151.

\bibitem[\protect\citeauthoryear{Hall and Johnstone}{Hall and
  Johnstone}{1992}]{HJ}
Hall, P. and I.~Johnstone (1992).
\newblock Empirical functional and efficient smoothing parameter selection.
\newblock {\em J. Roy. Statist. Soc. Ser. B\/}~{\em 54\/}(2), 475--530.
\newblock With discussion and a reply by the authors.

\bibitem[\protect\citeauthoryear{Hall and Marron}{Hall and
  Marron}{1987}]{H&M:Extent}
Hall, P. and J.~S. Marron (1987).
\newblock Extent to which least-squares cross-validation minimises integrated
  square error in nonparametric density estimation.
\newblock {\em Probab. Theory Related Fields\/}~{\em 74\/}(4), 567--581.

\bibitem[\protect\citeauthoryear{Hall and Marron}{Hall and
  Marron}{1991}]{HallMarron:LocMin}
Hall, P. and J.~S. Marron (1991).
\newblock Local minima in cross-validation functions.
\newblock {\em J. Roy. Statist. Soc. Ser. B\/}~{\em 53\/}(1), 245--252.

\bibitem[\protect\citeauthoryear{Hall and Schucany}{Hall and
  Schucany}{1989}]{Hall&Schucany:LocCV}
Hall, P. and W.~R. Schucany (1989).
\newblock A local cross-validation algorithm.
\newblock {\em Statist. Probab. Lett.\/}~{\em 8\/}(2), 109--117.

\bibitem[\protect\citeauthoryear{Hart and Vieu}{Hart and
  Vieu}{1990}]{Hart:Autocorrelation}
Hart, J.~D. and P.~Vieu (1990).
\newblock Data-driven bandwidth choice for density estimation based on
  dependent data.
\newblock {\em Ann. Statist.\/}~{\em 18\/}(2), 873--890.

\bibitem[\protect\citeauthoryear{Hart and Yi}{Hart and Yi}{1998}]{HartYi}
Hart, J.~D. and S.~Yi (1998).
\newblock One-sided cross-validation.
\newblock {\em J. Amer. Statist. Assoc.\/}~{\em 93\/}(442), 620--631.

\bibitem[\protect\citeauthoryear{Loader}{Loader}{1999}]{Loader:Classical}
Loader, C.~R. (1999).
\newblock Bandwidth selection: classical or plug-in?
\newblock {\em Ann. Statist.\/}~{\em 27\/}(2), 415--438.

\bibitem[\protect\citeauthoryear{Marron and Wand}{Marron and
  Wand}{1992}]{MarronWand:MISE}
Marron, J.~S. and M.~P. Wand (1992).
\newblock Exact mean integrated squared error.
\newblock {\em Ann. Statist.\/}~{\em 20\/}(2), 712--736.

\bibitem[\protect\citeauthoryear{Mielniczuk, Sarda, and Vieu}{Mielniczuk
  et~al.}{1989}]{Mielniczuk:LocCV}
Mielniczuk, J., P.~Sarda, and P.~Vieu (1989).
\newblock Local data-driven bandwidth choice for density estimation.
\newblock {\em J. Statist. Plann. Inference\/}~{\em 23\/}(1), 53--69.

\bibitem[\protect\citeauthoryear{Rudemo}{Rudemo}{1982}]{Rudemo:LSCV}
Rudemo, M. (1982).
\newblock Empirical choice of histograms and kernel density estimators.
\newblock {\em Scand. J. Statist.\/}~{\em 9\/}(2), 65--78.

\bibitem[\protect\citeauthoryear{Sain, Baggerly, and Scott}{Sain
  et~al.}{1994}]{Sainetal:cv}
Sain, S.~R., K.~A. Baggerly, and D.~W. Scott (1994).
\newblock Cross-validation of multivariate densities.
\newblock {\em J. Amer. Statist. Assoc.\/}~{\em 89\/}(427), 807--817.

\bibitem[\protect\citeauthoryear{Savchuk, Hart, and Sheather}{Savchuk
  et~al.}{2008}]{SavchukHartSheather:ICV}
Savchuk, O.~Y., J.~D. Hart, and S.~J. Sheather (2008).
\newblock Indirect cross-validation for density estimation.
\newblock {\em J. Amer. Statist. Assoc., submitted\/}.

\bibitem[\protect\citeauthoryear{Scott and Terrell}{Scott and
  Terrell}{1987}]{Scott:UCV}
Scott, D.~W. and G.~R. Terrell (1987).
\newblock Biased and unbiased cross-validation in density estimation.
\newblock {\em J. Amer. Statist. Assoc.\/}~{\em 82\/}(400), 1131--1146.

\bibitem[\protect\citeauthoryear{Sheather}{Sheather}{2004}]{Sheather:Density}
Sheather, S.~J. (2004).
\newblock Density estimation.
\newblock {\em Statist. Sci.\/}~{\em 19\/}(4), 588--597.

\bibitem[\protect\citeauthoryear{Sheather and Jones}{Sheather and
  Jones}{1991}]{Sheather:PI}
Sheather, S.~J. and M.~C. Jones (1991).
\newblock A reliable data-based bandwidth selection method for kernel density
  estimation.
\newblock {\em J. Roy. Statist. Soc. Ser. B\/}~{\em 53\/}(3), 683--690.

\bibitem[\protect\citeauthoryear{Stute}{Stute}{1992}]{Stute}
Stute, W. (1992).
\newblock Modified cross-validation in density estimation.
\newblock {\em J. Statist. Plann. Inference\/}~{\em 30\/}(3), 293--305.

\bibitem[\protect\citeauthoryear{Terrell}{Terrell}{1990}]{Terrell:OS}
Terrell, G.~R. (1990).
\newblock The maximal smoothing principle in density estimation.
\newblock {\em J. Amer. Statist. Assoc.\/}~{\em 85\/}(410), 470--477.

\bibitem[\protect\citeauthoryear{van Es}{van Es}{1992}]{vanEs}
van Es, B. (1992).
\newblock Asymptotics for least squares cross-validation bandwidths in
  nonsmooth cases.
\newblock {\em Ann. Statist.\/}~{\em 20\/}(3), 1647--1657.

\end{thebibliography}

\end{document}